\documentclass[]{aa} 
\usepackage{graphicx}
\usepackage{txfonts}
\usepackage{natbib}
\usepackage{url}
\usepackage{lscape}
\def\teff{$T\rm_{eff }$}
\def\kms{$\mathrm {km s}^{-1}$}
\def\kms{$\mathrm {km s}^{-1}$}
\newcommand{\halpha}{H$\alpha$}

\newcommand{\glog}{$\log$ g}
\newcommand{\fei}{\ion{Fe}{i}}
\newcommand{\feii}{\ion{Fe}{ii}}
\newcommand{\etab}{$\eta_{B}$}
\newcommand{\sli}{$^7$Li}
\newcommand{\hyd}{$^1$H}
\newcommand{\deu}{$^2$H}
\newcommand{\cobold}{{\tt CO$^5$BOLD}}
\newcommand{\linfor}{{\tt Linfor3D}}

\begin{document}
   \title{The metal-poor end of the Spite plateau}
   \subtitle{I: Stellar parameters, metallicities, and lithium abundances}

   \author{L. Sbordone\inst{1,2,3}
	\and
	P. Bonifacio\inst{1,2,4}
        \and
	E. Caffau\inst{2}
	\and
	H.-G. Ludwig\inst{1,2,5}	
	\and 
	N. T. Behara\inst{1,2,6}
	\and
	J.I. Gonz\'alez Hern\'andez\inst{1,2,7}
	\and
	M. Steffen\inst{8}
	\and
	R. Cayrel\inst{2}
	\and
	B. Freytag\inst{9}
	\and
	C. Van't Veer\inst{2}
	\and
	P. Molaro\inst{4}
	\and 
	B. Plez\inst{10}
	\and
	T. Sivarani\inst{11}
        \and
	M. Spite\inst{2}
	\and
	F. Spite\inst{2}
	\and
	T. C. Beers\inst{12}
	\and
	N.~Christlieb\inst{5}
        \and
	P. Fran\c cois\inst{2}
	\and
	V. Hill\inst{2,13}
	\fnmsep\thanks{Based on observations made with the ESO Very Large Telescope at Paranal Observatory, Chile (Programmes 076.A-0463 and 077.D-0299). Table 3 is fully available in electronic form
at the CDS via anonymous ftp to cdsarc.u-strasbg.fr (130.79.128.5) or via {\tt http://cdsweb.u-strasbg.fr/cgi-bin/qcat?J/A+A/}
}}

   \institute{
        %1
	CIFIST Marie Curie Excellence Team 
	\and
	%2
	GEPI, Observatoire de Paris, CNRS, Universit\'e Paris Diderot ; Place Jules Janssen, 92190 Meudon, France
	\and
	%3
	{ Max-Planck Institut f\"ur Astrophysik, Karl-Schwarzschild-Str. 1, 85741, Garching, Germany.\\
	\email{lsbordone@mpa-garching.mpg.de}}
	\and
	%4
	INAF - Osservatorio Astronomico di Trieste, via G. B. Tiepolo 11 34143 Trieste, Italy
	\and
	%5
	{Zentrum f\"ur Astronomie der Universit\"at Heidelberg, Landessternwarte, K\"onigstuhl 12, 69117 Heidelberg, Germany}
	\and
	%6
	{Institut d'Astronomie et d'Astrophysique, Universit\'e Libre de Bruxelles, CP226, boulevard du Triomphe, 1050 Bruxelles}
	\and
	%7	
	Dpto. de Astrof\'isica y Ciencias de la Atm\'osfera, Facultad de Ciencias F\'isicas, Universidad Complutense de Madrid, E-28040 Madrid, Spain
	%8
	\and
	Astrophysikalisches Institut Potsdam An der Sternwarte 16, D-14482 Potsdam, Germany
	\and
	%9
	Centre de Recherche Astrophysique de Lyon, UMR 5574: Universit\'e de Lyon, \'Ecole Normale Sup\'erieure de Lyon, 46 all\'ee d'Italie 69364 Lyon cedex 07. France
	\and
	%10
	Universit\'e Montpellier 2, CNRS, GRAAL, F-34095 Montpellier, France
	\and
	%11
	Indian Institute of Astrophysiscs, II Block, Koramangala, Bangalore 560 034, India
	\and
	%12
	Dept. if Physics \& Astronomy, and JINA: Joint Insrtitute for Nuclear Astrophysics, Michigan State University, E. Lansing, MI 48824, USA
	\and
	%13
	Cassiop\`ee - Observatoire de la Cote d'Azur, Boulevard de l'Observatoire, B.P. 4229 F-06304 NICE Cedex 4, France
        }

   \date{Received ...; accepted ...}

% \abstract{}{}{}{}{} 
% 5 {} token are mandatory
 
\abstract
% context heading (optional)
% {} leave it empty if necessary  
{The primordial nature of the Spite plateau is at odds with the WMAP satellite measurements, implying a primordial Li production at least three times higher than observed. It has also been suggested that A(Li) might exhibit a positive correlation with metallicity below [Fe/H]$\sim$-2.5. Previous samples studied comprised few stars below [Fe/H]=$-$3.}
% aims heading (mandatory)
{ We present VLT-UVES Li abundances of 28 Halo dwarf stars between [Fe/H]=$-$2.5 and $-$3.5, ten of which have [Fe/H]$<-$3.}
% methods heading (mandatory)
{We determined stellar parameters and abundances  using four different \teff\ scales. The direct infrared flux method was applied to infrared photometry. \halpha\ wings were fitted with two synthetic grids computed by means of 1D LTE atmosphere models, assuming two different self-broadening theories. A grid of \halpha\ profiles was finally computed by means of 3D hydrodynamical atmosphere models. The \ion{Li}{i} doublet at 670.8 nm has been used to measure A(Li) by means of 3D hydrodynamical NLTE spectral syntheses. An analytical fit of A(Li)$_{\rm 3D, NLTE}$ as a function of equivalent width, \teff, log g, and [Fe/H] has been derived and is made available.}
% results heading (mandatory)
{We confirm {{previous claims}} that A(Li) does not exhibit a plateau below [Fe/H]=$-$3. We detect a strong
positive correlation with [Fe/H] that is {{\em insensitive}} to
the choice of \teff\ estimator. From a linear fit, we infer a steep slope
of about 0.30 dex in A(Li) per dex in [Fe/H], which has a significance of 2-3 $\sigma$. The slopes derived using 
the four \teff\ estimators are consistent to within 1$\sigma$. 
A significant slope is also detected in the 
A(Li)--\teff\ plane, driven mainly by the coolest stars in the sample (\teff$<$6250), which appear to be Li-poor.
However, when we remove these stars the slope detected in the A(Li) -- [Fe/H] plane is not altered significantly.
When the full sample is considered, the scatter in A(Li)
increases by a factor of 2 towards lower metallicities, while the plateau appears
very thin above [Fe/H]=$-$2.8. At this metallicity, the plateau lies at
$\rm\left\langle{ A(Li)_{3D,NLTE}}\right\rangle=2.199\pm0.086$.} 
% conclusions heading (optional), leave it empty if necessary 
{The meltdown of the Spite plateau below [Fe/H]$\sim -3$ is established,
but its cause is unclear. If the primordial A(Li) were that derived from
standard BBN, it appears difficult to envision a single depletion phenomenon
producing a thin, metallicity independent plateau above [Fe/H]=$-$2.8, and a
highly scattered, metallicity dependent distribution below. That no
star below [Fe/H]=$-$3 lies above the plateau suggests that they formed at
plateau level and experienced subsequent depletion.}


\maketitle
%
%________________________________________________________________

\section{Introduction}

\citet{spite82Natur,spite82} first noted that metal-poor 
($-2.4$$\le$[Fe/H]$\le$$-1.4$), warm (5700 K $\le$ \teff $\le$ 6250 K), dwarf
stars exhibit a remarkably constant Li abundance, irrespective of metallicity and
effective temperature, and interpreted this {\it plateau} in Li abundance
(hereafter the {\it Spite plateau}) as being representative of the abundance of
Li synthesized during the primordial hot and dense phase of the Universe (Big
Bang, \citealt{wfh}; see \citealt{iocco08} for a review). Determining 
the lithium abundance in unevolved metal-poor stars has
since developed into an active research topic, because of its potential role as a cosmological
diagnostic. In the standard Big Bang nucleosynthesis (SBBN) scenario, \sli\ is
formed immediately after the Big Bang, together with 
\hyd, \deu, $^{3}$He, and $^{4}$He. \deu\ is formed first, and is 
subsequently required as a seed to form any heavier element (the so-called
``deuterium bottleneck''). The abundance of all the subsequent BBN products thus
depend on the equilibrium \deu\ abundance, which is determined by the \deu\
photodissociation reaction \deu($\gamma$,\hyd)\hyd. As a result, all the
abundances of BBN products ultimately depend on the primordial baryon/photon
ratio $\eta_{B}\equiv n_B / n_\gamma$ \citep{steigman01}, and can in
principle be employed to constrain this fundamental cosmological parameter. 

Following the cosmic microwave background anisotropy measurements of WMAP
\citep[e.g.][]{dunkley09}, \etab~ can be inferred from the value of the
baryonic density, $\Omega_B$, i. e. determining the primordial
abundance of BBN products is no longer the only means by which it is estimated. On the
other hand, the comparison between the two estimates remains of paramount
importance as a test of the reliability of the BBN theory, of our present
understanding of the subsequent chemical evolution of the elements involved,
and, in the case of Li, of our understanding of stellar atmospheres.

Among the available BBN products, \deu\ and \sli\ are the most reliable
\etab~ indicators. Being \deu\ never produced in stars, 
its abundance in a low-metallicity environment can be assumed to be
quite close to the cosmological value. In addition, its sensitivity to \etab\ is
monotonic and quite strong \citep[ (\deu/\hyd)$\propto \eta_B^{-1.6}$,
][]{steigman09}. On the other hand, \deu\ can be effectively measured only in
high-redshift, low-metallicity damped Lyman $\alpha$ (DLAs) or Lyman limit systems, for which
the observations are so challenging that only seven such
high quality measurements exist to date, which were all obtained after 10m-class
telescopes became available \citep[][]{pettini08}. The \etab~ value inferred from them
is in good agreement with that derived from WMAP \citep[][]{steigman09}.

In contrast, \sli\ can be measured with relative 
ease in the photospheres of warm, unevolved stars. The observations are typically
restricted to dwarfs, at least when one is interested in determining the
primordial Li abundance, because the fragile Li nucleus is destroyed by the
$^7$Li(p,$\alpha$)$^4$He reaction as soon as the temperature reaches 2.6 million
K. This implies that giants should not be considered, since their deep convective zones mix the surface
material with layers that exceed this temperature, and almost all Li is
rapidly destroyed. The ease with which \sli\ is destroyed has always constituted a challenge to
existing models of convection and diffusion in stellar atmospheres, which
predict a depletion of at least a factor of four relative to the
primordial abundance \citep{michaud84}. While one could infer that some
depletion might have occurred, it appeared impossible to obtain a {\em constant}
depletion over such a wide range of effective temperatures. The simplest solution
was to assume that no depletion was indeed taking place. This is in marked
contrast to the solar case, where the photospheric Li abundance is about
two dex lower than the meteoritic value. 

The original interpretation of the Spite plateau has been challenged in many ways in the
years since its discovery. Surely the most compelling challenge was the
independent measurement of \etab~ by the WMAP satellite, placing the
expected primordial Li abundance at A(Li)\footnote{A(Li)=$\log [N({\rm Li}) /
N({\rm H})]+12$}$_P$=2.65$^{+0.05}_{-0.06}$ \citep{steigman07}, or even higher,
A(Li)$_P$=2.72$\pm0.05$ when updated rates are taken into account for
the $^3$He($\alpha,\gamma$)$^7$Li reaction \citep[][]{cyburt08}. The highest
estimate of the Spite plateau does not exceed A(Li)=2.4, a more typical value
being A(Li)$\sim$ 2.2. The discrepancy can in principle be eliminated in two
ways, by either rejecting the standard BBN scenario \citep[for a review
see][]{iocco08}, or by assuming that some degree of Li depletion has occurred.
Two main mechanisms could again be invoked. Li could be subject to depletion
{\em before} the currently observed stars are formed \citep{piau06}, by means of
the reprocessing of the primordial gas in a first generation of massive, hot
stars. This phenomenon does not appear to be able to explain the entire
WMAP ~/~ Spite plateau gap, but, removing up to 0.3 dex of the discrepancy could
considerably reduce the problem. The maximum possible depletion is
nevertheless dependent on the initial mass function and lifetime of Pop. III
stars, as well as on the effectiveness of the mixing of their ejecta in the
interstellar medium, which are all poorly known. Alternatively, Li can be
depleted {\em within} the stars we currently observe, as a consequence of
phenomena within the envelope, such as diffusion, gravity waves, rotational
mixing, or any combination of these. As stated above, the negligible scatter,
and apparent lack of slope in the Spite plateau are observational constraints
that models of Li depletion have failed to reproduce. This could apparently be achieved 
by combining diffusion with some form of
turbulence at the bottom of the atmospheric convective zone 
\citep{richard05,korn06,korn07,piau08,lind09b}. 
Unfortunately, the effect of turbulence is introduced 
basically as a free parameter, and its tuning 
is made quite difficult by the subtlety of the effects 
expected on elements other than Li \citep[see sect. 6.4 in][]{bonifacio07}. 
Claims have been made \citep[e.g.][]{asplund06} that the lighter $^6$Li
isotope has been detected in the atmospheres of dwarf stars displaying Spite plateau $^7$Li abundances. 
These measurements are very difficult and sensitive to subtle details of the 
analysis \citep[][]{cayrel07}. If the detection of $^6$Li in EMP dwarf stars were to be confirmed, 
it would severely undermine any claim of a substantial atmospheric depletion 
of $^7$Li during the star's lifetime, since the $^6$Li is even more easily destroyed than $^7$Li.

One additional problem is constituted by repeated claims that
the Spite plateau might display a tilt towards lower Li abundances at lower
metallicities, on the order of 0.1-0.2 dex in A(Li) per dex in [Fe/H] 
\citep{ryan96,ryan99,boesgaard05,asplund06}, although other studies failed to confirm this \citep[e.g.][]{bonifacio1997}.
Roughly below [Fe/H]=$-$2.5, more and more stars appear to exhibit Li abundances below the
plateau level, while the scatter increases. 

The extreme case is possibly represented by the lithium abundance upper limit of
the hyper-iron-poor subgiant \object{HE 1327--2326} \citep[][and references
therein]{frebel08}, which should have A(Li)$\le$0.7 (from 1D analysis). The
interpretation of this result is not straightforward. 
Even rejecting the interpretation \citep{venn08} that this star might be a 
chemically-peculiar {\em evolved} object, the unusual photospherical composition of this star
has not yet found a satisfactory explanation. Were the composition of \object{HE 1327--2326}
to be indeed primordial, its lack of Li would support the
\citet{piau06} suggestion of a pollution by material cycled through massive
Pop. III stars. 

Adopting the \citet{piau06} hypothesis, one could then envision a scenario in
which partial pollution by this astrated material induces varying degrees of Li
``depletion'' in EMP stars according to how much this reprocessed gas is
available locally at the location and time of each stars' formation. A linear
fit to EMP stellar Li abundances would then naturally lead to an expected trend
in A(Li) with [Fe/H], whose slope would appear steeper the more the sample
is limited to low metallicities. An alternative explanation would be to postulate
that a Li ``over-depletion'' mechanism operates in the photospheres of the
most metal-poor stars, a mechanism that would not act uniformly in every star of
a given metallicity (possibly depending on
\teff\ or rotation speed or both), but would be more efficient at lower [Fe/H].
These stars would then begin with a Li abundance corresponding to the
Spite plateau, but most of them would then develop some degree of Li depletion.
This explanation would, at the same time, explain the apparent slope at low
metallicities and the increase in the scatter. It would also explain why, even
at very low metallicity, one still finds some stars lying on the Spite plateau.
A striking example of this is the EMP double-lined binary system
\object{CS 22876--032} \citep{gonzalez08}, in which, at [Fe/H]=$-$3.6, the primary lies 
on the Spite plateau, while the secondary has a Li abundance {\em lower} by
about 0.4 dex.

\begin{table}
\caption{Observations log for the 11 new targets. \label{obslog}}
\begin{center}
{\scriptsize
\begin{tabular}{llllll}
\hline
{\bf Star}              & {\bf Obs. date}      & {\bf MJD$^a$}   & {\bf Exp. time} & {\bf V$_{\mathrm rad}^b$} & {\bf S/N$^c$} \\
                        & {\bf (UT)     }      & {\bf (UT)   }   & {\bf sec.     } & {\bf km/s               } &               \\
\hline
\object{BS 17572--100}  & 21 Feb 2006          & 53787.05628641  & 3190				& 189                       & 191           \\
\\
\object{CS 22188--033}  & 7 Jul 2006           & 53923.31274319  & 3035            & 14                        & 107           \\
\\
\object{CS 22882--027}  & 6 Jul 2006           & 53922.31576391  & 4100            & 182                       & 78            \\
                        &                      & 53922.36486180  & 4100            & 182                       &               \\
\\
\object{CS 22950--173}  & 23 Apr 2006          & 53848.33688352  & 3600            & 69                        & 92            \\
\\
\object{CS 29491--084}  & 17 May 2006          & 53872.31418739  & 3600            & -8                        & 104           \\
\\
\object{CS 29514--007}  & 7 Jul 2006           & 53923.35103355  & 3600            & 41                        & 91            \\
\\
\object{CS 29516--028}  & 19 May 2006          & 53874.32992778  & 3600            & -179                      & 62            \\
                        &                      & 53874.37328577  & 3600            & -179                      &               \\
\\
\object{CS 30302--145}  & 23 Apr 2006          & 53848.28216715  & 2160            & 195                       & 70            \\
                        &                      & 53848.30823918  & 2160            & 195                       &               \\
\\
\object{CS 30344--070}  & 17 May 2006          & 53872.35925533  & 2160            & -141                      & 82            \\
                        & 27 May 2006          & 53882.31201609  & 2160            & -140                      &               \\
\\
\object{HE 0148--2611}  & 12 Jul 2006          & 53928.37446515  & 3600            & -227                      & 72            \\
\\
\object{HE 1413--1954}  & 21 Feb 2006          & 53787.27197427  & 3600            & -101                      & 52            \\
\hline
\multicolumn{6}{l}{\scriptsize $a$ Modified Julian date of observation start: MJD=JD-2400000.5}\\
\multicolumn{6}{l}{\scriptsize $b$ Rounded to the nearest km/s, barycentric correction applied}\\
\multicolumn{6}{l}{\scriptsize $c$ Near \ion{Li}{i} 670.8 nm doublet, if more than one spectrum has been used, S/N}\\
\multicolumn{6}{l}{\scriptsize ~~~is measured in the coadded spectrum.}\\
\end{tabular}}
\end{center}
\end{table}

\section{Observations and data reduction}

\begin{figure}
   \centering
   \includegraphics[width=9cm]{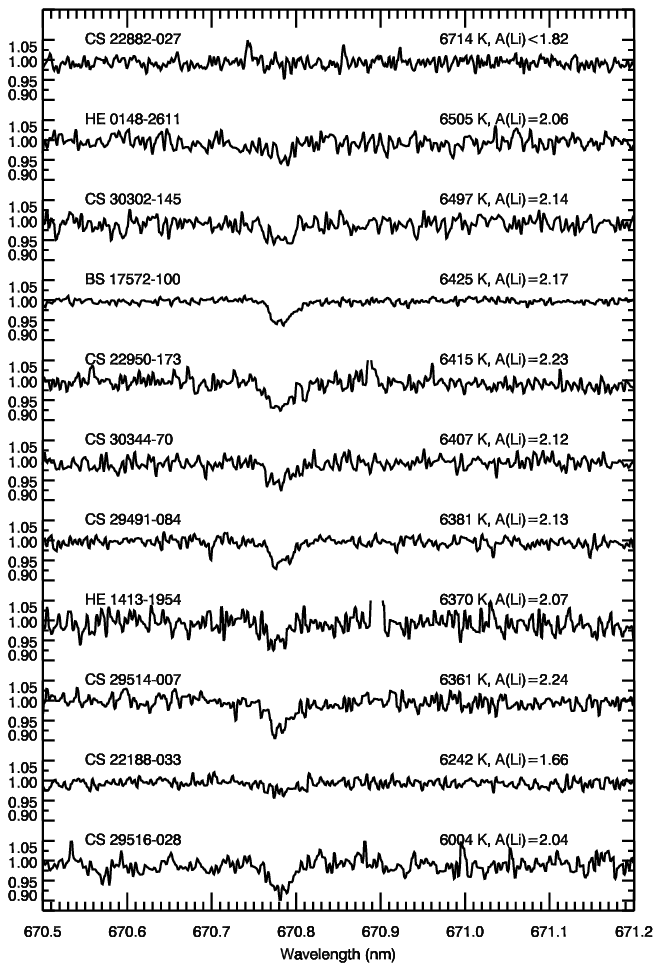}
      \caption{High-resolution spectra of the \ion{Li}{i} 670.79 nm doublet region for the 11 newly observed stars of the sample. For stars for which multiple spectra were available,
the coadded spectrum is shown. For the purpose of visualization, all the spectra have been shifted to zero radial velocity and normalized. \teff\ increases from bottom to top, 3D scale \teff\ and A(Li)$_{\rm 3D,NLTE}$ (see Sects. \ref{eff_temp} and \ref{li_abunds}) are listed for each star. The star \object{CS 22882--027} shows no detectable Li line, and the 3$\sigma$ upper limit to A(Li) is listed here.
%In stars \object{CS 22188-033} and \object{HE 0148-2611} the line is detected but not measurable. 
}
         \label{spectrasamples}
   \end{figure}

Our sample includes 11 main-sequence turnoff and dwarf stars selected from various sources (see Table \ref{coord_table}), along
with the sample already presented in \citet{bonifacio07}. The star \object{HE 0148--2611}
was previously analyzed \citep{cohen02,carretta02} but no Li measurement 
was ever performed. One star  (\object{HE 1413--1954}) was derived from the \citet{barklem05} sample. 
It, again, had no previous Li measurement. The remaining targets were drawn 
from the HK \citep{beers85,beers92,beers99} and Hamburg/ESO \citep{norbert08} 
surveys, and were never studied previously based on high-resolution spectra. They were observed by VLT-UVES \citep{dekker00} 
during programmes 076.A-0463(A) (P.I. {Lopez}, \object{HE 1413--1954} and 
\object{BS 17572--0100}) and 077.D-0299(A) (P.I. Bonifacio, remaining targets). 
{The observation log for the 11 new targets is in Table \ref{obslog}, where the final signal-to-noise ratio (S/N) 
around the \ion{Li}{i} 670.8 nm doublet is also indicated.} 
For the two stars observed during 076.A-0463(A), the observations were performed
by using VLT-UVES with the DIC1 dichroic and the 346nm + 580nm setting with a 1\farcs{0}
slit. These observations thus do not contain the 380-480 nm range, but for \object{HE
1413--1954} the HERES \citep{barklem05} data were available, which covered that
wavelength range. For the stars observed during 077.D-0299(A), we used DIC1 with
the 390nm + 580nm setting and 1\farcs{0} slit, thus providing coverage from 360nm to 750nm. 
{All the spectra have spectral resolution of R$\sim$40 000.}
The data were reduced using the standard UVES pipeline. In Fig. \ref{spectrasamples},
we show the \ion{Li}{i} 670.8\,nm line region for the 11 newly observed stars. 

The data were reduced and analyzed with the same procedures used in
\citet{bonifacio07}, to which the interested reader is referred for details of the
analysis and the associated uncertainties. An extract of the table listing the employed 
\ion{Fe}{i} and \ion{Fe}{ii} lines, in addition to associated atomic data, equivalent widths, and 
abundances in the 3D scale is available in Table \ref{line_by_line}. The full table is available in the online version.

\begin{table*}
\caption{Coordinates and optical and infrared photometry for the program stars.}
\label{coord_table}      
\centering          
{%\scriptsize
\begin{tabular}{l c c rrrrr}     % 7 columns 
\hline
{\bf Star} & {\boldmath{$\alpha$}} & {\boldmath{$\delta$}} & {\bf V} & {\bf J} & {\bf H} & {\bf K} & {\bf E(B-V)}\\
%           & h & m & s & $^\circ$ & ' & '' & & & \\
\hline
 \object{BS 16023--046} & 14h 00m 54s.6 & +22$^\circ$ 46' 48''   & 14.17 & 13.24 & 13.02 & 12.96 & 0.01801\\
 \object{BS 17570--063} & 00h 20m 36s.1 & +23$^\circ$ 47' 38''   & 14.51 & 13.47 & 13.17 & 13.07 & 0.03949\\
 \object{BS 17572--100} & 09h 28m 55s.3 & $-$05$^\circ$ 21' 36'' & 12.17 & 11.28 & 11.02 & 10.95 & 0.03727\\
 \object{CS 22177--009} & 04h 07m 40s.5 & $-$25$^\circ$ 02' 40'' & 14.27 & 13.25 & 12.96 & 13.03 & 0.04407\\
 \object{CS 22188--033} & 00h 51m 25s.9 & $-$38$^\circ$ 12' 18'' & 13.20 & 12.16 & 11.91 & 11.90 & 0.01315\\
 \object{CS 22882--027} & 00h 38m 09s.7 & $-$31$^\circ$ 47' 54'' & 15.11 &  --   &  --   &  --   & --     \\
 \object{CS 22888--031} & 23h 11m 32s.4 & $-$35$^\circ$ 26' 43'' & 14.90 & 13.91 & 13.75 & 13.65 & 0.01417\\
 \object{CS 22948--093} & 21h 50m 31s.5 & $-$41$^\circ$ 07' 49'' & 15.18 & 14.29 & 13.98 & 14.00 & 0.01576\\
 \object{CS 22950--173} & 20h 35m 31s.2 & $-$15$^\circ$ 53' 30'' & 14.04 & 12.98 & 12.70 & 12.66 & 0.04551\\
 \object{CS 22953--037} & 01h 25m 06s.8 & $-$59$^\circ$ 15' 58'' & 13.64 & 12.68 & 12.44 & 12.46 & 0.02796\\
 \object{CS 22965--054} & 22h 06m 30s.0 & $-$02$^\circ$ 32' 39'' & 15.10 & 13.86 & 13.58 & 13.45 & 0.13321\\
 \object{CS 22966--011} & 23h 35m 06s.6 & $-$30$^\circ$ 22' 53'' & 14.55 & 13.54 & 13.23 & 13.27 & 0.01391\\
 \object{CS 29491--084} & 22h 28m 49s.5 & $-$28$^\circ$ 57' 03'' & 13.48 & 12.52 & 12.25 & 12.20 & 0.01367\\
 \object{CS 29499--060} & 23h 53m 40s.2 & $-$26$^\circ$ 58' 44'' & 13.03 & 12.10 & 11.85 & 11.86 & 0.02027\\
 \object{CS 29506--007} & 21h 20m 28s.6 & $-$20$^\circ$ 46' 24'' & 14.18 & 13.17 & 12.93 & 12.87 & 0.04547\\
 \object{CS 29506--090} & 21h 30m 28s.9 & $-$22$^\circ$ 10' 41'' & 14.33 & 13.34 & 13.10 & 13.07 & 0.04547\\
 \object{CS 29514--007} & 01h 06m 40s.6 & $-$24$^\circ$ 58' 41'' & 13.97 & 12.96 & 12.67 & 12.66 & 0.02375\\
 \object{CS 29516--028} & 22h 25m 40s.3 & +05$^\circ$ 37' 40''   & 15.02 & 13.63 & 13.29 & 13.15 & 0.12816\\
 \object{CS 29518--020} & 01h 12m 12s.9 & $-$31$^\circ$ 00' 06'' & 14.00 & 13.06 & 12.76 & 12.74 & 0.02241\\
 \object{CS 29518--043} & 01h 18m 38s.2 & $-$30$^\circ$ 41' 02'' & 14.57 & 13.64 & 13.35 & 13.37 & 0.02030\\
 \object{CS 29527--015} & 00h 29m 10s.5 & $-$19$^\circ$ 10' 07'' & 14.25 & 13.29 & 13.08 & 13.05 & 0.02213\\
 \object{CS 30301--024} & 15h 08m 29s.7 & $-$00$^\circ$ 36' 02'' & 12.95 & 11.93 & 11.67 & 11.67 & 0.06527\\
 \object{CS 30302--145} & 19h 40m 52s.2 & $-$48$^\circ$ 39' 19'' & 14.46 & 13.48 & 13.23 & 13.26 & 0.05343\\
 \object{CS 30339--069} & 00h 30m 15s.9 & $-$35$^\circ$ 56' 51'' & 14.75 & 13.77 & 13.52 & 13.45 & 0.00904\\
 \object{CS 30344--070} & 22h 47m 23s.2 & $-$35$^\circ$ 32' 44'' & 14.43 & 13.53 & 13.26 & 13.27 & 0.01305\\
 \object{CS 31061--032} & 02h 38m 43s.1 & +03$^\circ$ 19' 03''   & 13.90 & 12.87 & 12.62 & 12.61 & 0.03727\\
 \object{HE 0148--2611} & 01h 50m 59s.5 & $-$25$^\circ$ 57' 02'' & 14.45 & 13.55 & 13.30 & 13.30 & 0.01362\\
 \object{HE 1413--1954} & 14h 16m 04s.7 & $-$20$^\circ$ 08' 54'' & 15.23 & 14.19 & 13.97 & 13.89 & 0.08681\\
 \object{LP 815--43}    & 20h 38m 13s.3 & $-$20$^\circ$ 26' 11'' & 10.91 &  9.96 &  9.71 &  9.65 & 0.04514\\
\hline
\end{tabular}}
\end{table*}

\section{Atmosphere models and spectrosynthesis programs}

\subsection{1D LTE models and spectrosynthesis}

Various one-dimensional (1D) local thermodynamical equilibrium (LTE) atmosphere
models were employed in the present study. F. Castelli's grid of fluxes
computed using {\tt ATLAS 9} (\citealt{castelli03}\footnote{grid available at
\url{http://wwwuser.oat.ts.astro.it/castelli/}}) was used in
the infrared flux temperature determination (see Sect. \ref{irfm}). A second {\tt
ATLAS 9} model grid \citep{kurucz05,sbordone04,sbordone05} was computed
with an ad hoc mixing length parameter in producing the
\halpha-wing
profiles used to determine \teff\ (see Sect. \ref{hafit}). {It has been shown \citep[][]{fuhrmann93,vantveer96} that employing in ATLAS a mixing length
parameter of $l/H_p=0.5$ provides the best fit to Balmer lines profiles in the Sun, while the value $l/H_p=1.25$ generally better reproduces
more closely the solar flux, and thus, is usually employed in ``general purpose'' models.} To compute these 
profiles, we employed a modified version of R. L. Kurucz's code {\tt
BALMER}\footnote{The original version is available online at
\url{http://kurucz.harvard.edu/}}, which was capable of handling different line-broadening
theories. Finally, we employed {\tt OSMARCS} atmosphere models
\citep{gustafsson75,plez92,edvardsson93,asplund97,gustafsson03} and the {\tt
turbospectrum} spectral synthesis code \citep{alvarez98} to determine
\ion{Fe}{i} and \ion{Fe}{ii} abundances, gravity, microturbulence, and Li
abundances. Hydrostatic monodimensional {\tt LHD} models \citep[see][]{CL07,caffau07}
were used in determining the 1D NLTE corrections (see Sect. \ref{3dnlte}).

\subsection{3D hydrodynamical models and spectrosynthesis}

\label{3dmodels}

Time-dependent, hydrodynamical 3D stellar atmosphere models computed with
\cobold\ \citep{freytag02,wedemeyer04} as part of the CIFIST model grid (\citealt{ludwig09b}\footnote{see \url{http://cifist.obspm.fr/}})
 were employed to produce grids
of \halpha-wing profiles for \teff\ estimation \citep[see Sect. \ref{hafit},
and ][]{ludwig09,behara09}. 

\section{Effective temperature}
\label{eff_temp}

Effective temperature (\teff) is the most crucial stellar atmosphere parameter
influencing Li abundance determination, Li abundances derived from the
\ion{Li}{i} 670.75nm line being sensitive to \teff\ at the level of about 0.03 dex
for each 50~K variation in \teff. Unfortunately, a precise determination of
stellar effective temperatures is generally difficult to achieve. For F/G
dwarf and subgiant stars such as those studied here, \teff\ is routinely
estimated either from photometric calibrations \citep[e.g.,][]{alonso00,alonso01}
or by fitting the wings of \halpha\ with a grid of synthetic profiles of
varying \teff. 

\begin{table*}
\caption{An extract from the line-by-line \ion{Fe}{i} and \ion{Fe}{ii} abundance table. The full table is available online. \label{line_by_line}}
\begin{tabular}{lllllllll}
\hline
{\bf Star}             & {\bf Ion}    & ${\boldmath \lambda}$& ${\boldmath \log gf}$ & {\bf EW}  & ${\boldmath \epsilon}$ & ${\boldmath \epsilon}$ & ${\boldmath \epsilon}$ & ${\boldmath \epsilon}$ \\
                       &              & {\bf (nm) }          &                       & {\bf pm }& {\bf  BA        }       & {\bf  ALI       }       & {\bf  IRFM      }      & {\bf  3D    }         \\
\hline
\object{BS 16023--046} &  \ion{Fe}{i} &  340.1519 & -2.059 &  14.60 &  4.85  & 5.01 &  5.04  & {\bf 4.89} \\
                       &  \ion{Fe}{i} &  340.7460 & -0.020 &  28.60 &  4.38  & 4.51 &  4.54  & {\bf 4.41} \\
                       &  \ion{Fe}{i} &  342.7119 & -0.098 &  29.40 &  4.50  & 4.47 &  4.61  & {\bf 4.63} \\
                       &  \ion{Fe}{i} &  344.0989 & -0.958 &  76.50 &  4.77  & 4.92 &  4.95  & {\bf 4.76} \\
...                    &              &           &        &        &       \\
\hline
\end{tabular}
\end{table*}

Both methods are plagued by specific accuracy issues. Photometric calibrations,
or the infrared flux method (IRFM), are mainly sensitive to the accuracy of the
photometry available, to the details of the calibration process, and to 
uncertainties in the interstellar reddening estimates. On the other hand, \halpha\ fitting 
is mainly sensitive to both the uncertainty in the continuum normalization across the
broad line wings, and the choice of the broadening theory applied in the line synthesis (see
Sect. \ref{hafit}).

In this paper, we considered four temperature estimators:
\begin{itemize}
  \item temperatures derived from \halpha-wing fitting, using 1D atmosphere
models and spectrosynthesis, self broadening being treated according to \citet{barklem00,barklem00b} and
Stark broadening according to \citet{stehle99} (we will henceforth refer to these temperatures as ``BA
temperatures'', or the ``BA temperature scale'');
  \item same as BA, but using the \citet{ali66} self-broadening theory 
(ALI temperatures);
  \item temperatures derived from \halpha-wing fitting, using 3D model atmospheres 
and spectrosynthesis, \citet{barklem00,barklem00b} self broadening and
  \citet{stehle99} Stark broadening (3D temperatures, for all \halpha\ derived
  temperatures see Sect. \ref{hafit}); 
\item temperature derived with the infrared
  flux method \citep[see Sect. \ref{irfm} as well as][IRFM temperatures]{gonzalez09}.
\end{itemize}

\subsection{Fitting of the \halpha\ wings}

\label{hafit} Temperature scales based on \halpha-wing fitting are affected
by both observational and theoretical issues. Most high-resolution spectrographs
use echelle gratings operating in high orders, which exhibit a steep blaze
function. The continuum placement is thus sensitive to the accuracy with which
the shape of the grating blaze function can be estimated. Such uncertainties are
irrelevant when studying narrow lines observed at high resolution, but are
important when a broad feature such as \halpha\ is considered. More generally,
the precision of continuum placement and of the determination of the \halpha\-
wing shape are affected by noise as well as by the possible presence of weak
unrecognized features (less of a problem for metal-poor stars). Among these,
the blaze function shape likely introduces the largest uncertainty.

   \begin{figure}
   \centering
   \includegraphics[width=9cm]{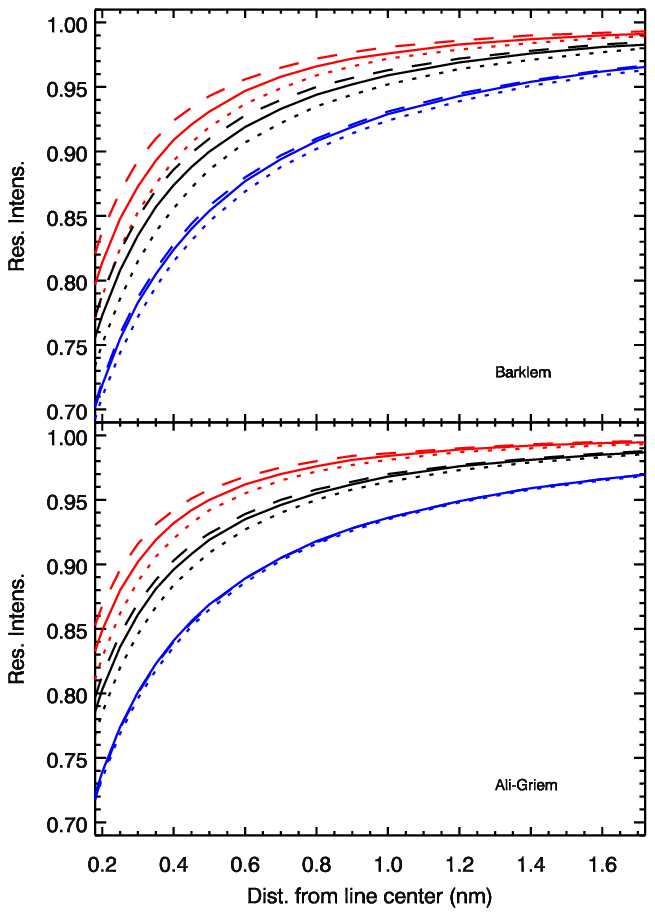}
      \caption{\halpha\  red wing profiles in the wavelength range significant for the fit. In each panel, red profiles (upper ones) are for \teff=5400 K, black \
      profiles (middle ones) are for \teff=6000 K, and blue profiles (lower) for \teff=6600 K. For each temperature, dashed profiles are for \glog=3.5, solid profiles 
      for \glog=4.0, and dotted profiles are for \glog=4.5. All profiles assume [Fe/H]=$-$3. Upper panel shows profiles for BA temperatures, lower panel for ALI temperatures.}
         \label{tsens}
   \end{figure}

On the theoretical side, the uncertainties are due both to the atmosphere model
structure and to the physics employed in the \halpha\ synthesis. \halpha-wing
self broadening can be treated with different theories, most notably those of
\citet{ali66}, \citet{barklem00,barklem00b}, and \citet{allard08}. As a
general rule, \citet{ali66} theory leads to a significantly {\em lower}
broadening coefficient with respect to the ones derived from \citet{barklem00,
barklem00b} and \citet{allard08}. A significantly higher \teff\ is required
to reproduce a given observed profile when employing the \citet{ali66} theory with
respect to the other theories. With the typical parameters of the stars in our
sample, and using our fitting procedure, employing \citet{ali66} self broadening
leads to derived \teff\ estimates that are higher by about 150-200 K (a
difference of about 0.1 dex in Li abundance) with respect to those derived by
using the \citet{barklem00,barklem00b} theory. The theory by \citet{allard08}, on
the other hand, leads to \teff\ within a few tens of K of the \teff\ estimates
obtained when using the \citet{barklem00,barklem00b} theory. We thus restricted
ourselves to using the self-broadening theory of \citet{barklem00,barklem00b} (in BA
and 3D temperatures) and the \citet{ali66} self-broadening theory (ALI
temperatures). 

The \halpha-fitting temperatures exhibit a
significant gravity sensitivity. \citet{barklem02} already reported estimates of this 
sensitivity for relatively metal-poor models (down to [Fe/H]=$-$2). The effect
is always in the sense of higher gravity leading to broader profiles, and
appears generally stronger at lower metallicities, at lower temperatures, and
for the BA profiles compared to the ALI profiles. In Fig. \ref{tsens}, we plot examples of
profiles for the ALI and BA cases. Profiles are plotted for \teff=5400,
6000, and 6600~K (higher temperatures generate broader profiles). A metallicity
of [Fe/H]=$-$3 is used. For each temperature, we plot the profile for \glog=3.5,
4, and 4.5. As can be noted, the part of the wing closest to the core appears to
be more strongly affected than other parts. It is clearly seen that the gravity
sensitivity of the BA profiles is roughly twice as large as in the ALI case.  In both the BA and ALI scales, the
gravity effect becomes quickly negligible as \teff\ increases above 6500~K. In the most
deviant cases ([Fe/H]$<$ $-$3, \teff$<$6000 K, BA profiles), a difference of 0.5 dex
in \glog\ leads to roughly a 200~K difference in \teff.

\begin{table*}
\caption{Atmosphere parameters for the program stars using the different temperature
 estimators. Parameters for \object{CS 22882--027} were derived only for the 3D temperature scale.}             
\label{params}      
\centering          
{\scriptsize
\begin{tabular}{l r c c l r c c l r c c l r c c l}     % 7 columns 
\hline
{\bf Star} & {\bf T}{\boldmath $_{\rm eff}$} & {\boldmath $\log{g}$} & {\boldmath $\xi$} & {\bf [Fe/H]} & {\bf T}{\boldmath $_{\rm eff}$} & {\boldmath $\log{g}$} & {\boldmath $\xi$} & {\bf [Fe/H]} & {\bf T}{\boldmath $_{\rm eff}$} & {\boldmath $\log{g}$} & {\boldmath $\xi$} & {\bf [Fe/H]} & {\bf T}{\boldmath $_{\rm eff}$} & {\boldmath $\log{g}$} & {\boldmath $\xi$} & {\bf [Fe/H]}\\
 & K & CGS & \kms &   & K & CGS & \kms &   & K & CGS & \kms &   & K & CGS & \kms &   \\
\hline
  & \multicolumn{4}{c}{\bf BA } & \multicolumn{4}{c}{\bf ALI } & \multicolumn{4}{c}{\bf IRFM } & \multicolumn{4}{c}{\bf 3D } \\
\hline
 \object{BS 16023--046} & 6324 & 4.30 & 1.4 & $-$2.97 & 6527 & 4.60 & 1.4 & $-$2.84 & 6560 & 4.60 & 1.4 & $-$2.82 & 6401 & 4.50 & 1.4 & $-$2.94 \\ 
 \object{BS 17570--063} & 6078 & 4.50 & 0.6 & $-$3.05 & 6404 & 4.80 & 0.7 & $-$2.79 & 6315 & 4.70 & 0.7 & $-$2.86 & 6237 & 4.70 & 0.6 & $-$2.92 \\ 
 \object{BS 17572--100} & 6371 & 4.00 & 1.6 & $-$2.75 & 6504 & 4.40 & 1.5 & $-$2.62 & 6689 & 4.70 & 1.5 & $-$2.52 & 6425 & 4.30 & 1.5 & $-$2.72 \\ 
 \object{CS 22177--009} & 6177 & 4.30 & 1.3 & $-$3.17 & 6415 & 4.70 & 1.3 & $-$2.99 & 6479 & 4.70 & 1.3 & $-$2.96 & 6284 & 4.50 & 1.2 & $-$3.08 \\ 
 \object{CS 22188--033} & 6129 & 4.40 & 1.4 & $-$3.03 & 6411 & 4.90 & 1.3 & $-$2.85 & 6281 & 4.50 & 1.4 & $-$2.98 & 6242 & 4.70 & 1.2 & $-$2.97 \\ 
 \object{CS 22882--027} &  --  &  --  & --  &     --  &  --  &  --  & --  &   --    &  --  &  --  & --  &   --    & 6714 & 4.70 & 1.4 & $-$2.40 \\
 \object{CS 22888--031} & 5925 & 4.50 & 0.7 & $-$3.47 & 6304 & 5.10 & 0.7 & $-$3.18 & 6480 & 5.20 & 1.0 & $-$3.07 & 6090 & 4.90 & 0.4 & $-$3.33 \\ 
 \object{CS 22948--093} & 6365 & 4.25 & 1.3 & $-$3.31 & 6551 & 4.50 & 1.3 & $-$3.15 & 6577 & 4.70 & 1.2 & $-$3.18 & 6450 & 4.40 & 1.3 & $-$3.24 \\ 
 \object{CS 22950--173} & 6335 & 4.20 & 1.4 & $-$2.78 & 6506 & 4.50 & 1.4 & $-$2.61 & 6353 & 4.20 & 1.4 & $-$2.73 & 6415 & 4.40 & 1.4 & $-$2.69 \\ 
 \object{CS 22953--037} & 6325 & 4.25 & 1.4 & $-$2.91 & 6515 & 4.50 & 1.4 & $-$2.75 & 6557 & 4.45 & 1.4 & $-$2.76 & 6416 & 4.40 & 1.4 & $-$2.84 \\ 
 \object{CS 22965--054} & 6245 & 4.00 & 1.5 & $-$2.90 & 6398 & 4.20 & 1.5 & $-$2.78 & 6417 & 4.20 & 1.5 & $-$2.79 & 6312 & 4.10 & 1.4 & $-$2.86 \\ 
 \object{CS 22966--011} & 6049 & 4.40 & 1.1 & $-$3.22 & 6345 & 4.90 & 1.1 & $-$2.96 & 6302 & 4.80 & 1.1 & $-$3.01 & 6166 & 4.70 & 1.0 & $-$3.09 \\ 
 \object{CS 29491--084} & 6285 & 4.00 & 1.7 & $-$3.04 & 6453 & 4.20 & 1.8 & $-$2.90 & 6425 & 4.20 & 1.8 & $-$2.94 & 6381 & 4.10 & 1.8 & $-$2.97 \\ 
 \object{CS 29499--060} & 6349 & 4.10 & 1.5 & $-$2.66 & 6493 & 4.40 & 1.4 & $-$2.56 & 6560 & 4.50 & 1.5 & $-$2.56 & 6428 & 4.30 & 1.5 & $-$2.62 \\ 
 \object{CS 29506--007} & 6285 & 4.20 & 1.6 & $-$2.88 & 6478 & 4.40 & 1.7 & $-$2.70 & 6515 & 4.40 & 1.7 & $-$2.71 & 6397 & 4.30 & 1.7 & $-$2.81 \\ 
 \object{CS 29506--090} & 6287 & 4.20 & 1.4 & $-$2.83 & 6480 & 4.55 & 1.4 & $-$2.67 & 6557 & 4.45 & 1.5 & $-$2.63 & 6367 & 4.30 & 1.4 & $-$2.77 \\ 
 \object{CS 29514--007} & 6281 & 4.10 & 1.5 & $-$2.80 & 6448 & 4.40 & 1.5 & $-$2.66 & 6351 & 4.30 & 1.4 & $-$2.79 & 6361 & 4.30 & 1.5 & $-$2.76 \\ 
 \object{CS 29516--028} & 5839 & 4.40 & 1.2 & $-$3.52 & 6198 & 5.00 & 1.2 & $-$3.19 & 5994 & 4.70 & 1.2 & $-$3.39 & 6004 & 4.90 & 0.9 & $-$3.33 \\ 
 \object{CS 29518--020} & 6127 & 4.30 & 1.8 & $-$2.86 & 6368 & 4.80 & 1.8 & $-$2.67 & 6471 & 4.90 & 1.9 & $-$2.60 & 6213 & 4.60 & 1.8 & $-$2.79 \\ 
 \object{CS 29518--043} & 6376 & 4.25 & 1.3 & $-$3.25 & 6566 & 4.40 & 1.4 & $-$3.10 & 6537 & 4.25 & 1.4 & $-$3.16 & 6489 & 4.30 & 1.4 & $-$3.17 \\ 
 \object{CS 29527--015} & 6276 & 4.00 & 1.6 & $-$3.53 & 6426 & 4.40 & 1.6 & $-$3.37 & 6578 & 4.50 & 1.7 & $-$3.31 & 6325 & 4.30 & 1.6 & $-$3.49 \\ 
 \object{CS 30301--024} & 6375 & 4.00 & 1.6 & $-$2.71 & 6494 & 4.50 & 1.6 & $-$2.60 & 6581 & 4.50 & 1.6 & $-$2.60 & 6400 & 4.30 & 1.5 & $-$2.69 \\ 
 \object{CS 30302--145} & 6403 & 4.30 & 1.8 & $-$3.02 & 6597 & 4.50 & 1.8 & $-$2.88 & 6645 & 4.50 & 1.9 & $-$2.88 & 6497 & 4.40 & 1.8 & $-$2.94 \\ 
 \object{CS 30339--069} & 6253 & 4.00 & 1.4 & $-$3.09 & 6402 & 4.40 & 1.4 & $-$2.93 & 6375 & 4.40 & 1.3 & $-$2.98 & 6304 & 4.30 & 1.3 & $-$3.04 \\ 
 \object{CS 30344--070} & 6302 & 4.10 & 1.6 & $-$3.02 & 6477 & 4.30 & 1.7 & $-$2.85 & 6568 & 4.40 & 1.8 & $-$2.82 & 6407 & 4.20 & 1.7 & $-$2.92 \\ 
 \object{CS 31061--032} & 6369 & 4.25 & 1.4 & $-$2.62 & 6555 & 4.50 & 1.5 & $-$2.48 & 6405 & 4.25 & 1.4 & $-$2.58 & 6466 & 4.40 & 1.5 & $-$2.56 \\ 
 \object{HE 0148--2611} & 6400 & 4.10 & 1.5 & $-$3.18 & 6565 & 4.30 & 1.6 & $-$3.06 & 6606 & 4.20 & 1.6 & $-$3.07 & 6505 & 4.20 & 1.6 & $-$3.12 \\ 
 \object{HE 1413--1954} & 6302 & 3.80 & 1.7 & $-$3.50 & 6448 & 4.10 & 1.7 & $-$3.39 & 6716 & 4.40 & 1.8 & $-$3.22 & 6370 & 4.00 & 1.7 & $-$3.47 \\ 
 \object{LP 815--43}    & 6453 & 3.80 & 1.7 & $-$2.88 & 6579 & 4.10 & 1.7 & $-$2.81 & 6630 & 4.10 & 1.7 & $-$2.77 & 6578 & 4.00 & 1.7 & $-$2.80 \\
\hline
\end{tabular}}
\end{table*}

\begin{figure}
   \centering
   \includegraphics[width=9cm]{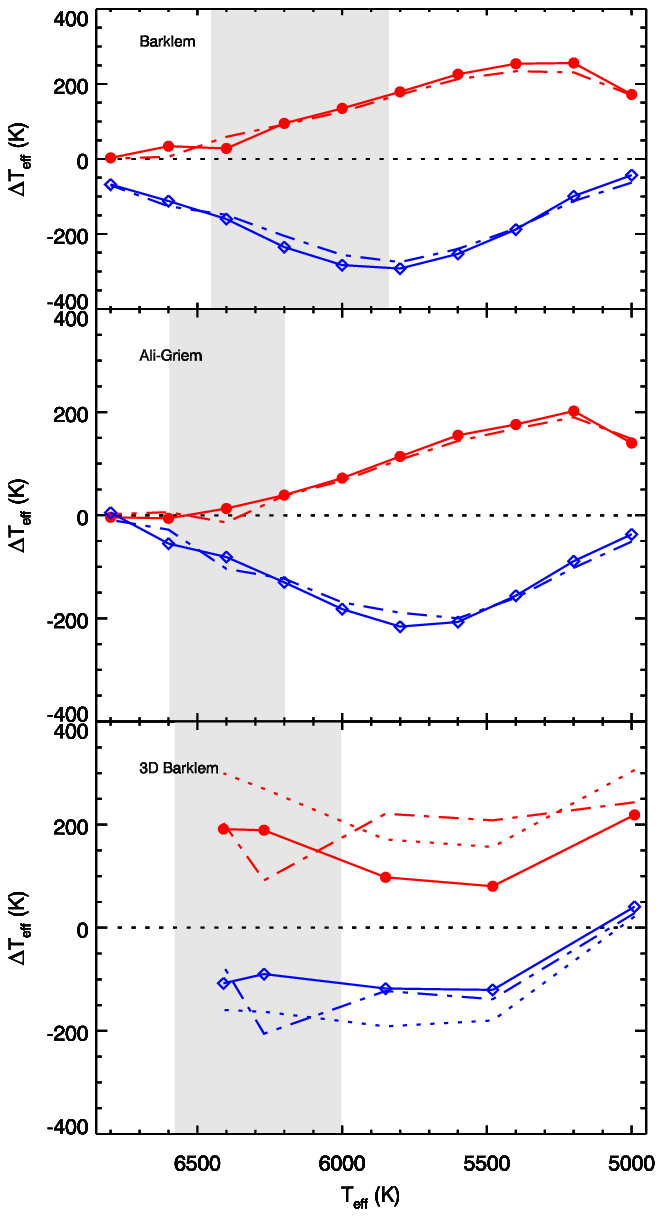}
      \caption{\halpha\ theoretical profiles for models with \glog=4,
      [Fe/H]=$-$3 (solid lines) and \glog=4, [Fe/H]=$-$2.5 (dot-dashed lines)
      have been fitted with the same procedure used for the program stars, but
      using a \glog=3.5 grid (red lines with filled circles) and a \glog=4.5
      grid (blue lines with open diamonds). Here we plot the temperature
      difference (\teff(\glog=3.5/4.5)$-$\teff(\glog=4.0) ), against the
      ``real'' effective temperature of the profile. The upper panel shows BA profiles,
      the middle panel ALI profiles, and lower panel 3D profiles. The gray shaded areas
      indicates the temperature ranges for the program stars in each \teff\
      scale.} \label{thadif} \end{figure}

Thus, the shape of the \halpha\ profile varies in different ways when varying
gravity and temperature. As a consequence, the use of an incorrect value of gravity will
always affect the temperature estimate, but the {\em size} of the effect will
depend on the details of how the actual fitting is performed. To provide some
insight into what the effect is when employing our specific fitting procedure, we
fed the fitting program with \glog=4, and [Fe/H]=$-$3 and $-$2.5 theoretical
\halpha\ profiles, and derived the temperature by assuming that \glog=3.5, 4.0, and
4.5. When fitting profiles of \glog=4.0 by means of profiles of \glog=3.5,
which are narrower at each temperature, we obtain a {\em higher} temperature
estimate than we would if we were to use the proper gravity. The opposite effect occurs 
when using the broader \glog=4.5 profiles. In Fig. \ref{thadif}, we plot
these temperature differences versus the true \teff\ of the profiles.
Differences are computed in the sense $\Delta$\teff$=$\teff(\glog=3.5 or 4.5)
$-$\teff(\glog=4.0). Red lines with filled circles correspond to fits with
\glog=3.5 profiles, blue ones with open diamonds to fits with \glog=4.5 profiles.
The solid lines correspond to [Fe/H]=$-$3.0 profiles (both fitted and fitting),
while the dot-dashed line corresponds to [Fe/H]=$-$2.5. For the parameter
space covered, and when adopting our fitting procedure, {\em underestimating} the 
fitting-grid gravity by 0.5 dex leads to an {\em overestimate} of the temperature by
as much as 250~K in the BA case, and 200~K in the ALI case. This underestimate reaches a 
maximum around 5200-5300 K, decreasing on both sides, and fading away on the
hot side, near \teff=6500 K. By {\em overestimating} the fitting-grid gravity, one
{\em underestimates} \teff\ by as much as 300~K in the BA case and 200~K in the
ALI case. The shape of the curve is similar, but the point of maximum
sensitivity occurs between 5800 and 6000~K. There is a hint that the effect
decreases mildly at [Fe/H]= $-2.5$, although higher metallicities have not been explored.
   
Since we estimate surface gravity from the \fei-\feii~ ionization equilibrium, the derived
gravity is temperature sensitive, so that the two estimations need to be
iterated to convergence. As a general rule, we stopped iterating when \teff\
variations became lower than 50~K, which typically required not more than 3
iterations, starting from an initial guess of \glog=4. 

   \begin{figure}
   \centering
   \includegraphics[width=9cm]{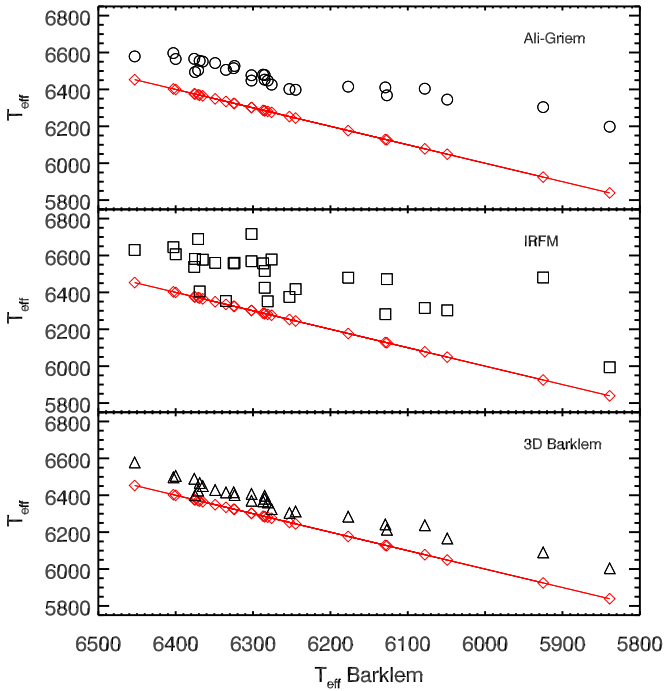}
      \caption{Effective temperatures for different estimators, plotted against
	      BA temperature for the program stars. Top to bottom: ALI, IRFM and
	      3D temperatures. The red line represents the one-to-one relation
	      (hence the line of BA temperatures). } \label{tvst} \end{figure}

A very mild metallicity sensitivity is also present in the \halpha\-based temperature determination,
never surpassing some tens of K for a 0.5 dex of variation in [Fe/H]. The actual
iteration of the temperature determination with the other atmosphere parameters
was performed differently for the BA and ALI cases on one side, and for the 3D
case on the other side:

\begin{itemize}
\item In the BA and ALI case, once the gravity and metallicity were determined 
with one temperature estimate, the \halpha\ profile grid was interpolated to that
gravity value, while the nearest grid step was chosen in metallicity, without
interpolation. The small metallicity step of the grid (0.25 dex), as well as the
very mild sensitivity of \halpha\ to metallicity, made this choice sufficiently
precise.
\item In the 3D case, the computation of both model atmospheres and spectral
 synthesis is very time consuming, and only the atmosphere model grids for \glog=4 and \glog=4.5 with
 [Fe/H]=$-$3.0 were sufficiently extended at the time of
 the analysis. We thus fitted the observed \halpha\ lines to these two grids,
 deriving, for each star, effective temperatures corresponding to the two
 assumed gravities. We determined gravity and metallicity, then derived a new
 \teff\ estimate by linearly interpolating between the \teff(\glog=4.0) and
 \teff(\glog=4.5) at the estimated gravity. The procedure was then iterated but,
 for most stars, the same convergence criterion applied to the 1D case
 ($\Delta$T$<$50~K) was found to be too stringent, since the parameters for most stars
 ended up oscillating between two sets corresponding to \teff\ estimates 
 that were about 60~K apart.
 This can probably be attributed to the use of a more coarse grid in the 3D case.
\end{itemize} 

   \begin{figure} \centering \includegraphics[width=9cm]{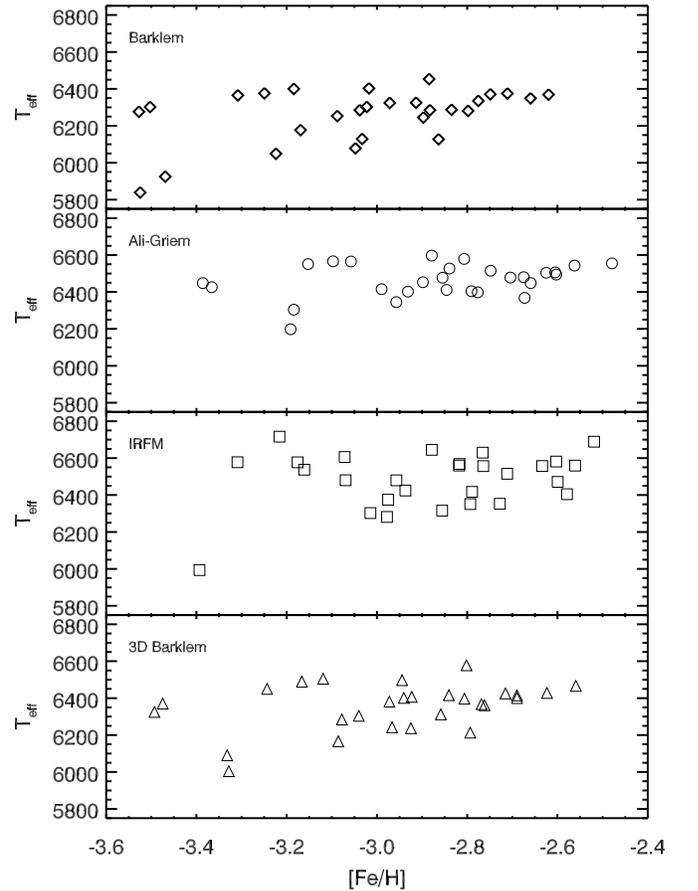}
   \caption{Effective temperatures for different estimators, plotted against
   [Fe/H] (as derived using the temperature in the panel). Top to bottom: BA
   temperatures, then ALI, IRFM and 3D.} \label{tvsmet} \end{figure}

\subsection{IRFM temperature estimation}
\label{irfm}

Originally introduced by \citet{BS77}, and later improved by \citet{B80}, who
removed an unnecessary iteration \citep[see][and references therein]{B90}, the
infrared flux method (IRFM) relies on the ratio of the flux at a near-infrared 
(NIR) wavelength or in a NIR band, to the bolometric flux. This ratio can
also be derived from model atmospheres, and the effective temperature determined
by finding the effective temperature of the model that reproduces the observed
ratio. \citet{gonzalez09} presented a new implementation of the method making
use of 2MASS photometry \citep{2MASS}, and also provided a calibration of
bolometric fluxes with colors $(V-J)$, $(V-H)$, and $(V-K_s)$, where V is in the
Johnson system and the NIR magnitudes are in the 2MASS system. We applied the
IRFM in exactly the way described by \citet{gonzalez09}, deriving the bolometric fluxes
as the average of those estimated from the three visible-NIR colors. All
magnitudes and colors used in the IRFM must be corrected for reddening. To do so,
we used the reddening maps of \citet{schlegel}, corrected as described
in \citet{B00}. All our program stars are sufficiently distant that they lie outside
the dust layer, so that the full reddening derived from the maps should be applied.
The adopted reddenings are provided in Table \ref{coord_table}.
The star \object{CS 22882--027} does not appear in the 2MASS catalog, thus we
could not derive its IRFM temperature.

\section{Gravity, microturbulence and metallicity}

The {\tt FITLINE} code was employed to measure the equivalent widths of
the \fei~ and \feii~
lines. Although up to $\sim$120 \fei\ lines were available, only four
\feii~ lines were strong enough to be used. {For each temperature scale, gravity was then derived by
enforcing \fei-\feii\ ionization equilibrium. For the \halpha-based scales, gravity was used with metallicity
(\fei~ abundance) to iterate the \teff\ estimation (see Sect. \ref{hafit}). }

{For each temeperature scale, }microturbulence was determined by ensuring that the weak and strong 
\ion{Fe}{i} lines provide the same abundance. The final parameters and the derived metallicity for
each temperature estimator are presented in Table \ref{params}, detailed \ion{Fe}{i} and \ion{Fe}{ii} 
abundances are listed in Table \ref{iron_abu}. 
Final \teff\ values for the ALI, IRFM, and 3D temperature scale are plotted against the BA scale in 
Fig. \ref{tvst}, and against the respective value of [Fe/H] in Fig.\ref{tvsmet}.

\section{Lithium abundance determination}
\label{li_abunds}

We determined Li equivalent widths in a similar fashion to
\citet{bonifacio07}. Synthetic line profiles were fitted to the observed
profile, and the equivalent width (EW) determined from the fitted synthetic
profile. The EW errors listed in Table \ref{tabliabu} were obtained by
means of Monte Carlo simulations, in which Poisson noise was added to a
synthetic spectrum to ensure that it had the same S/N as the observed spectrum. The
Li abundance was determined by iteratively computing synthetic spectra of
the Li doublet until the synthetic EW matched the observed EW to better than
1\%. The adopted atomic data were unchanged with respect to \citet{bonifacio07}, and
took account of hyperfine structure and isotopic components (a solar Li isotopic
ratio was assumed). We henceforth refer to these abundances as ``1D Li
abundances'' since Li abundances were derived using 1D atmosphere model and
spectrosynthesis codes. One should avoid confusion with the 3D temperature
scale, which indicates only that 3D effects have been taken into
account in determining the \halpha-wing fitting temperature.

In addition, we determined, {\em for the 3D temperature scale only}, what we
refer to as ``3D NLTE Li abundances''. As described in Sect. \ref{3dnlte}, a grid
of time-dependent 3D NLTE spectrosyntheses have been produced for the
\ion{Li}{i} 670.8 nm doublet, and used to independently determine Li abundances
from the measured EW. 

\begin{table*}
\caption{\ion{Fe}{i} and \ion{Fe}{ii} mean abundances, as well as their associated $\sigma$ for the four temperature scales. \label{iron_abu}}
\centering
{\scriptsize
\begin{tabular}{lllllllllllllllll}
\hline
{\bf Star}               & \ion{Fe}{i} & $\sigma$ & \ion{Fe}{ii} & $\sigma$ & \ion{Fe}{i} & $\sigma$ & \ion{Fe}{ii} & $\sigma$ & \ion{Fe}{i} & $\sigma$ & \ion{Fe}{ii} & $\sigma$ & \ion{Fe}{i} & $\sigma$ & \ion{Fe}{ii} & $\sigma$ \\
 & \multicolumn{4}{c}{\bf BA } & \multicolumn{4}{c}{\bf ALI } & \multicolumn{4}{c}{\bf IRFM } & \multicolumn{4}{c}{\bf 3D } \\
\hline
 \object{BS 16023--046}  & 4.53 & 0.108 & 4.50 & 0.135 &  4.66 & 0.113 & 4.64 & 0.135 &  4.68 & 0.115 & 4.64 & 0.135 &  4.56 & 0.108 & 4.58 & 0.136 \\
 \object{BS 17570--063}  & 4.45 & 0.136 & 4.51 & 0.062 &  4.71 & 0.156 & 4.66 & 0.062 &  4.64 & 0.151 & 4.62 & 0.061 &  4.58 & 0.144 & 4.61 & 0.062 \\
 \object{BS 17572--100}  & 4.75 & 0.124 & 4.71 & 0.091 &  4.88 & 0.158 & 4.89 & 0.085 &  4.98 & 0.172 & 5.02 & 0.086 &  4.78 & 0.152 & 4.84 & 0.086 \\
 \object{CS 22177--009}  & 4.33 & 0.101 & 4.37 & 0.123 &  4.51 & 0.121 & 4.54 & 0.126 &  4.54 & 0.114 & 4.55 & 0.127 &  4.42 & 0.116 & 4.46 & 0.123 \\
 \object{CS 22188--033}  & 4.47 & 0.104 & 4.44 & 0.030 &  4.65 & 0.123 & 4.66 & 0.032 &  4.52 & 0.097 & 4.49 & 0.032 &  4.53 & 0.110 & 4.57 & 0.027 \\
 \object{CS 22888--031}  & 4.03 & 0.148 & 4.04 & 0.125 &  4.32 & 0.164 & 4.30 & 0.132 &  4.43 & 0.173 & 4.35 & 0.137 &  4.17 & 0.154 & 4.21 & 0.126 \\
 \object{CS 22948--093}  & 4.19 & 0.133 & 4.20 & 0.144 &  4.35 & 0.132 & 4.31 & 0.146 &  4.32 & 0.129 & 4.38 & 0.145 &  4.26 & 0.134 & 4.26 & 0.146 \\
 \object{CS 22950--173}  & 4.72 & 0.121 & 4.76 & 0.111 &  4.89 & 0.126 & 4.90 & 0.110 &  4.77 & 0.103 & 4.77 & 0.110 &  4.81 & 0.124 & 4.85 & 0.111 \\
 \object{CS 22953--037}  & 4.59 & 0.126 & 4.62 & 0.124 &  4.75 & 0.145 & 4.74 & 0.125 &  4.74 & 0.136 & 4.72 & 0.125 &  4.66 & 0.128 & 4.69 & 0.125 \\
 \object{CS 22965--054}  & 4.60 & 0.136 & 4.62 & 0.089 &  4.72 & 0.141 & 4.71 & 0.091 &  4.71 & 0.140 & 4.71 & 0.091 &  4.64 & 0.138 & 4.67 & 0.083 \\
 \object{CS 22966--011}  & 4.28 & 0.106 & 4.30 & 0.120 &  4.54 & 0.131 & 4.53 & 0.121 &  4.49 & 0.119 & 4.48 & 0.121 &  4.41 & 0.118 & 4.44 & 0.119 \\
 \object{CS 29491--084}  & 4.46 & 0.130 & 4.49 & 0.119 &  4.60 & 0.134 & 4.58 & 0.119 &  4.56 & 0.133 & 4.57 & 0.119 &  4.53 & 0.132 & 4.53 & 0.119 \\
 \object{CS 29499--060}  & 4.84 & 0.114 & 4.82 & 0.099 &  4.94 & 0.127 & 4.96 & 0.097 &  4.94 & 0.113 & 4.98 & 0.100 &  4.88 & 0.120 & 4.89 & 0.100 \\
 \object{CS 29506--007}  & 4.62 & 0.116 & 4.67 & 0.183 &  4.80 & 0.125 & 4.77 & 0.186 &  4.79 & 0.124 & 4.77 & 0.186 &  4.69 & 0.118 & 4.72 & 0.186 \\
 \object{CS 29506--090}  & 4.67 & 0.110 & 4.71 & 0.109 &  4.83 & 0.109 & 4.82 & 0.113 &  4.87 & 0.122 & 4.82 & 0.113 &  4.73 & 0.114 & 4.76 & 0.109 \\
 \object{CS 29514--007}  & 4.70 & 0.130 & 4.66 & 0.084 &  4.84 & 0.136 & 4.80 & 0.087 &  4.71 & 0.132 & 4.75 & 0.078 &  4.74 & 0.133 & 4.74 & 0.086 \\
 \object{CS 29516--028}  & 3.98 & 0.135 & 3.97 & 0.184 &  4.31 & 0.151 & 4.24 & 0.183 &  4.11 & 0.141 & 4.10 & 0.184 &  4.17 & 0.150 & 4.18 & 0.184 \\
 \object{CS 29518--020}  & 4.64 & 0.092 & 4.62 & 0.138 &  4.83 & 0.115 & 4.83 & 0.139 &  4.90 & 0.125 & 4.87 & 0.141 &  4.71 & 0.100 & 4.74 & 0.139 \\
 \object{CS 29518--043}  & 4.25 & 0.126 & 4.28 & 0.129 &  4.40 & 0.128 & 4.36 & 0.133 &  4.34 & 0.124 & 4.30 & 0.133 &  4.33 & 0.125 & 4.31 & 0.133 \\
 \object{CS 29527--015}  & 3.97 & 0.143 & 3.98 & 0.179 &  4.13 & 0.143 & 4.15 & 0.179 &  4.19 & 0.142 & 4.19 & 0.179 &  4.01 & 0.145 & 4.09 & 0.179 \\
 \object{CS 30301--024}  & 4.79 & 0.118 & 4.73 & 0.106 &  4.90 & 0.119 & 4.92 & 0.108 &  4.90 & 0.119 & 4.92 & 0.108 &  4.81 & 0.125 & 4.85 & 0.102 \\
 \object{CS 30302--145}  & 4.48 & 0.160 & 4.48 & 0.102 &  4.62 & 0.160 & 4.58 & 0.107 &  4.62 & 0.160 & 4.58 & 0.107 &  4.56 & 0.163 & 4.53 & 0.103 \\
 \object{CS 30339--069}  & 4.41 & 0.188 & 4.39 & 0.098 &  4.57 & 0.179 & 4.56 & 0.101 &  4.52 & 0.177 & 4.55 & 0.096 &  4.46 & 0.189 & 4.51 & 0.096 \\
 \object{CS 30344--070}  & 4.48 & 0.121 & 4.51 & 0.104 &  4.65 & 0.123 & 4.61 & 0.101 &  4.68 & 0.125 & 4.65 & 0.101 &  4.58 & 0.122 & 4.56 & 0.102 \\
 \object{CS 31061--032}  & 4.88 & 0.137 & 4.88 & 0.062 &  5.02 & 0.132 & 4.98 & 0.067 &  4.92 & 0.123 & 4.89 & 0.062 &  4.94 & 0.125 & 4.93 & 0.068 \\
 \object{HE 0148--2611}  & 4.32 & 0.112 & 4.31 & 0.048 &  4.44 & 0.118 & 4.41 & 0.051 &  4.43 & 0.116 & 4.41 & 0.051 &  4.38 & 0.117 & 4.36 & 0.050 \\
 \object{HE 1413--1954}  & 4.00 & 0.118 & 4.01 & 0.167 &  4.11 & 0.118 & 4.13 & 0.167 &  4.28 & 0.118 & 4.27 & 0.168 &  4.03 & 0.120 & 4.09 & 0.168 \\
 \object{LP 815--43}     & 4.62 & 0.088 & 4.58 & 0.051 &  4.69 & 0.089 & 4.71 & 0.052 &  4.73 & 0.091 & 4.71 & 0.052 &  4.70 & 0.090 & 4.67 & 0.052 \\
\hline
\end{tabular}}
\end{table*}

The uncertainties in the Li abundance measurements were largely dominated by the
uncertainty in the temperature estimation. For further details, the reader is
referred to \citet{bonifacio07}. For the purpose of our analysis, a constant
uncertainty of $\sigma_{\rm A(Li)}=0.09$ was assumed.

\subsection{3D and NLTE corrections}
\label{3dnlte}

We originally planned to determine the effects of both atmosphere hydrodynamics and any 
departure from LTE in a consistent manner, and thus computed a set of
time-dependent 3D NLTE syntheses of the Li doublet over a grid of suitable 3D
models, to construct a set of curves of growth (COG) for the doublet
EW. Details of the computation of the 3D NLTE lithium doublet synthesis are
covered in Appendix \ref{3dnlte_calc}.

\begin{table*}
\caption{\ion{Li}{i} 670.8 nm EW and errors, and lithium abundances using the 
different parameter sets. For the BA, ALI, and IRFM temperature scales, we list
1D LTE and 1D NLTE A(Li). For the 3D temperature scale, we list 1D LTE and NLTE, 
as well as 3D NLTE A(Li).}
\label{tabliabu}      
\centering          
{\scriptsize
\begin{tabular}{l c c c c c c c c c c   c  }     
\hline
{\bf Star} &{\bf EW} & {\bf error} & {\bf A(Li) }&{\bf A(Li) }  &{\bf A(Li) }&{\bf A(Li) }   &{\bf A(Li) }&{\bf A(Li) }     &{\bf A(Li) }&{\bf A(Li) }&{\bf A(Li) }\\
           &  pm     &  pm         & ($a$)       & ($b$)        & ($a$)      &($b$)          & ($a$)      & ($b$)           & ($a$)      &($b$)       & ($c$)      \\
           &         &             & \multicolumn{2}{c}{\bf BA }& \multicolumn{2}{c}{\bf ALI}& \multicolumn{2}{c}{\bf IRFM} &\multicolumn{3}{c}{\bf 3D} \\
\hline
 
\object{BS 16023--046}  & 1.93 & 0.06 & 2.145 & 2.138 & 2.271 & 2.257 & 2.292 & 2.278 & 2.193 & 2.181 & {\bf 2.179}  \\
\object{BS 17570--063}  & 1.76 & 0.04 & 1.930 & 1.928 & 2.148 & 2.132 & 2.091 & 2.077 & 2.038 & 2.025 & {\bf 2.029}  \\
\object{BS 17572--100}  & 1.81 & 0.04 & 2.152 & 2.149 & 2.232 & 2.219 & 2.340 & 2.329 & 2.184 & 2.174 & {\bf 2.166}  \\
\object{CS 22177--009}  & 2.42 & 0.03 & 2.153 & 2.153 & 2.309 & 2.293 & 2.339 & 2.323 & 2.224 & 2.214 & {\bf 2.209}  \\
\object{CS 22188--033}  & 0.78 & 0.05 & 1.577 & 1.577 & 1.750 & 1.735 & 1.665 & 1.657 & 1.648 & 1.636 & {\bf 1.665}  \\
\object{CS 22888--031}  & 1.87 & 0.04 & 1.846 & 1.851 & 2.104 & 2.081 & 2.214 & 2.199 & 1.961 & 1.940 & {\bf 1.976}  \\
\object{CS 22948--093}  & 1.19 & 0.06 & 1.935 & 1.930 & 2.047 & 2.034 & 2.051 & 2.039 & 1.988 & 1.977 & {\bf 1.989}  \\
\object{CS 22950--173}  & 2.11 & 0.09 & 2.199 & 2.193 & 2.307 & 2.293 & 2.212 & 2.205 & 2.250 & 2.238 & {\bf 2.230}  \\
\object{CS 22953--037}  & 1.95 & 0.03 & 2.151 & 2.145 & 2.272 & 2.258 & 2.278 & 2.265 & 2.211 & 2.200 & {\bf 2.194}  \\
\object{CS 22965--054}  & 2.21 & 0.06 & 2.161 & 2.164 & 2.263 & 2.255 & 2.277 & 2.268 & 2.206 & 2.203 & {\bf 2.185}  \\
\object{CS 22966--011}  & 1.37 & 0.05 & 1.788 & 1.792 & 1.983 & 1.966 & 1.945 & 1.929 & 1.867 & 1.855 & {\bf 1.869}  \\
\object{CS 29491--084}  & 1.77 & 0.07 & 2.080 & 2.083 & 2.189 & 2.180 & 2.171 & 2.163 & 2.142 & 2.138 & {\bf 2.128}  \\
\object{CS 29499--060}  & 2.07 & 0.06 & 2.201 & 2.196 & 2.322 & 2.309 & 2.311 & 2.297 & 2.250 & 2.239 & {\bf 2.231}  \\
\object{CS 29506--007}  & 2.05 & 0.04 & 2.149 & 2.146 & 2.275 & 2.262 & 2.298 & 2.285 & 2.223 & 2.214 & {\bf 2.205}  \\
\object{CS 29506--090}  & 1.85 & 0.05 & 2.102 & 2.098 & 2.225 & 2.211 & 2.274 & 2.261 & 2.156 & 2.147 & {\bf 2.137}  \\
\object{CS 29514--007}  & 2.33 & 0.09 & 2.211 & 2.209 & 2.320 & 2.307 & 2.231 & 2.222 & 2.263 & 2.254 & {\bf 2.241}  \\
\object{CS 29516--028}  & 2.41 & 0.12 & 1.904 & 1.919 & 2.159 & 2.136 & 2.000 & 1.990 & 2.025 & 2.004 & {\bf 2.035}  \\
\object{CS 29518--020}  & 2.10 & 0.11 & 2.052 & 2.053 & 2.211 & 2.195 & 2.276 & 2.260 & 2.109 & 2.099 & {\bf 2.091}  \\
\object{CS 29518--043}  & 1.72 & 0.11 & 2.121 & 2.115 & 2.238 & 2.225 & 2.204 & 2.193 & 2.193 & 2.182 & {\bf 2.183}  \\
\object{CS 29527--015}  & 1.86 & 0.06 & 2.091 & 2.098 & 2.188 & 2.177 & 2.272 & 2.259 & 2.121 & 2.116 & {\bf 2.113}  \\
\object{CS 30301--024}  & 1.77 & 0.06 & 2.143 & 2.140 & 2.213 & 2.200 & 2.250 & 2.237 & 2.157 & 2.148 & {\bf 2.139}  \\
\object{CS 30302--145}  & 1.54 & 0.11 & 2.086 & 2.077 & 2.203 & 2.190 & 2.215 & 2.203 & 2.143 & 2.131 & {\bf 2.138}  \\
\object{CS 30339--069}  & 2.04 & 0.11 & 2.125 & 2.130 & 2.223 & 2.212 & 2.204 & 2.194 & 2.157 & 2.151 & {\bf 2.140}  \\
\object{CS 30344--070}  & 1.68 & 0.09 & 2.064 & 2.063 & 2.177 & 2.166 & 2.231 & 2.218 & 2.135 & 2.128 & {\bf 2.120}  \\
\object{CS 31061--032}  & 2.10 & 0.06 & 2.221 & 2.213 & 2.336 & 2.322 & 2.250 & 2.241 & 2.282 & 2.269 & {\bf 2.262}  \\
\object{HE 0148--2611}  & 1.29 & 0.09 & 2.000 & 1.996 & 2.100 & 2.089 & 2.113 & 2.102 & 2.065 & 2.056 & {\bf 2.063}  \\
\object{HE 1413--1954}  & 1.59 & 0.14 & 2.035 & 2.047 & 2.129 & 2.123 & 2.285 & 2.274 & 2.079 & 2.080 & {\bf 2.070}  \\
\object{LP 815--43}     & 1.89 & 0.06 & 2.229 & 2.228 & 2.302 & 2.292 & 2.334 & 2.324 & 2.304 & 2.296 & {\bf 2.296}  \\
\hline                                                                                                    
\multispan{6}{{\scriptsize $a$ 1D LTE value.} \hfill}\\                                                   
\multispan{6}{{\scriptsize $b$ 1D value with NLTE correction applied.} \hfill}\\                          
\multispan{6}{{\scriptsize $c$ 3D NLTE calculation} \hfill}\\                                             
\end{tabular}}                                                                                            
\end{table*} 

The model parameters covered by the COG grid are listed in Table \ref{3dgrid}.
Once the COG grid was computed, we decided to also derive 3D NLTE lithium abundances
directly, by identifying the EW-to-abundance relation that most closely fitted the computed
values, and applying it to our observed EW. This was accomplished by either
interpolating in the \teff, \glog, [Fe/H], and EW grid (and possibly
extrapolating out of it), or by determining a best fitting analytical function
in the form A(Li)=f(\teff, \glog, [Fe/H], EW) and applying it to the observed
parameters and lithium doublet EW. We pursued both of these approaches.

   \begin{figure}
   \centering
   \includegraphics[width=9cm]{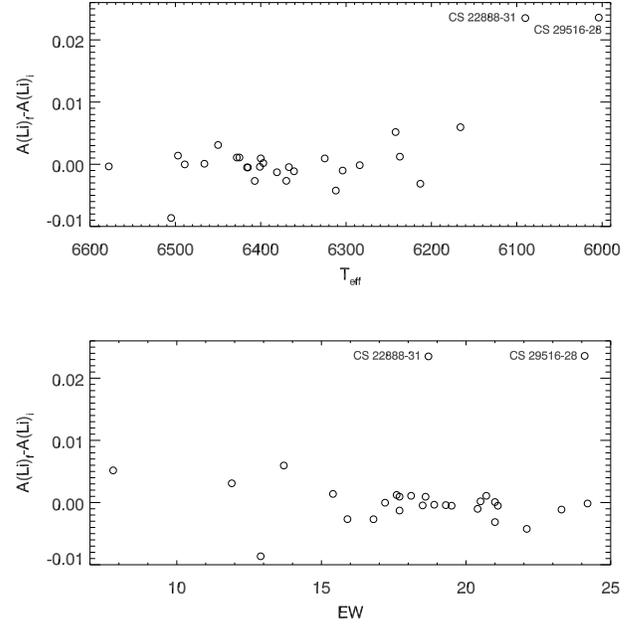}
      \caption{Difference between A(Li)$_f$ and A(Li)$_i$ plotted against \teff\
	      and the Li doublet EW. The two ``outliers'' are labeled. }
	      \label{intp_vs_fit_1} \end{figure}

We found the functional-fit method to be the superior of the two, both because of its
higher accuracy and for greater ease of use. Its functional-form
approach condenses the 3D NLTE abundance determination into a formula that can be
hard-coded into any program, eliminating the need to carry over the true grid
of computed points. Details on the fit calculation, as well as the chosen
functional form and coefficients, are available in Appendix
\ref{fitappendix}\footnote{The corresponding IDL functions are also available
on-line at \url{http://mygepi.obspm.fr/~sbordone/fitting.html} or by email request to the authors.}.

As mentioned above, we also performed an interpolation over the COG grid. The main
problem in producing a suitable interpolation lies is that the grid is
non-rectangular, which has two different causes. First, in the \cobold\
hydrodynamical models \teff\ is not set {\it a priori}, rather, the
entropy of the material entering through the bottom of the computational box is
the fixed quantity.
The true \teff\ is determined after snapshot selection, and usually
varies across an interval of $\pm$100~K centered on the desired value for the models we employed. As a
consequence, it is impossible to build a grid of \cobold\ models with exactly
the same temperature but, for example, different metallicity. Secondly, varying the stellar
parameters naturally alters the relationship between A(Li) and EW, so that the
range of EW in the COG corresponding to interesting values of A(Li) will vary from model
to model. 

   \begin{figure}
   \centering
   \includegraphics[width=9cm]{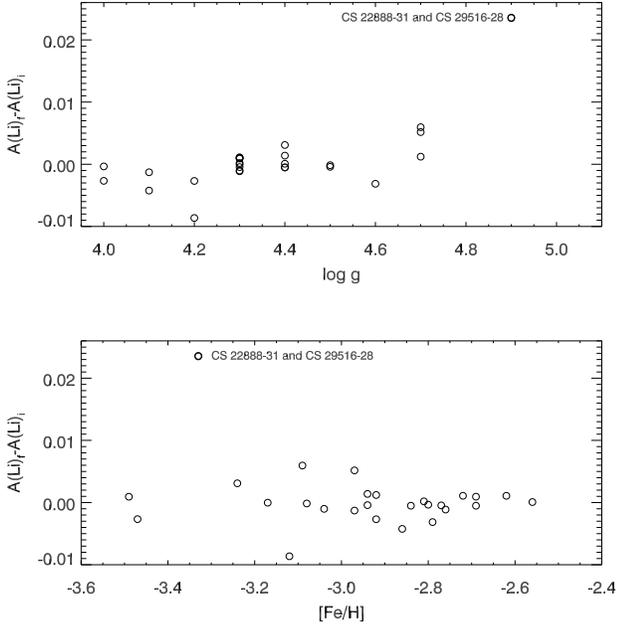}
      \caption{Same as in Fig. \ref{intp_vs_fit_1} but now plotting against log
	      g and [Fe/H]. } \label{intp_vs_fit_2} \end{figure}

To simplify the task, we took advantage of the limited sensitivity of the
670.8nm Li doublet to both gravity and metallicity. Thus, we decided to
assume [Fe/H]=$-$3 throughout the interpolation, and to avoid interpolating in
gravity by always choosing the closest value to the derived gravity 
between \glog=4 and \glog=4.5. This
choice was also justified by the limited extension in both parameters of our
sample. This reduced the problem to interpolating in an irregularly spaced
two-dimensional grid in \teff\ and EW. Delaunay triangulation\footnote{Delaunay
triangulation is a method of triangulation of a set P of points in a plane
defined as the triangulation for which no element of P lies within the
circumcircle of each triangle, except for the triangle vertexes. It is often
used to model surfaces that are sampled on irregular grids (e.g., elevations in
geography). The built-in IDL functions {\tt triangulate} and {\tt trigrid} have been
used to produce the triangulation and the interpolation based on it.} and
quintic polynomial interpolation were then used to derive A(Li). {\

Figures \ref{intp_vs_fit_1} and \ref{intp_vs_fit_2} show the difference
between A(Li) as determined by means of the analytical fit (A(Li)$_f$) and by
interpolation (A(Li)$_i$), respectively, plotted against relevant quantities, on the 3D
temperature scale. Most stars show an excellent concordance between the two
methods, but two outliers exist, \object{CS 29516--028} and \object{CS
22888--031}. These two stars have the lowest temperatures among all the stars in
the sample (these temperatures are still within the computed grid). However, they 
also have the highest gravity in the sample, which requires
extrapolation, since the grid has a limiting gravity of \glog\ of 4.5. While the
functional fit is indeed extrapolated, the simplified interpolation assumes log
g =4.5 in this instance; the discrepancy between the two methods does
not however exceed 0.023 dex in A(Li), which is negligible for our purpose. All the
remaining stars exhibit discrepancies not exceeding 0.01 dex.

   \begin{figure}
   \centering
   \includegraphics[width=9cm]{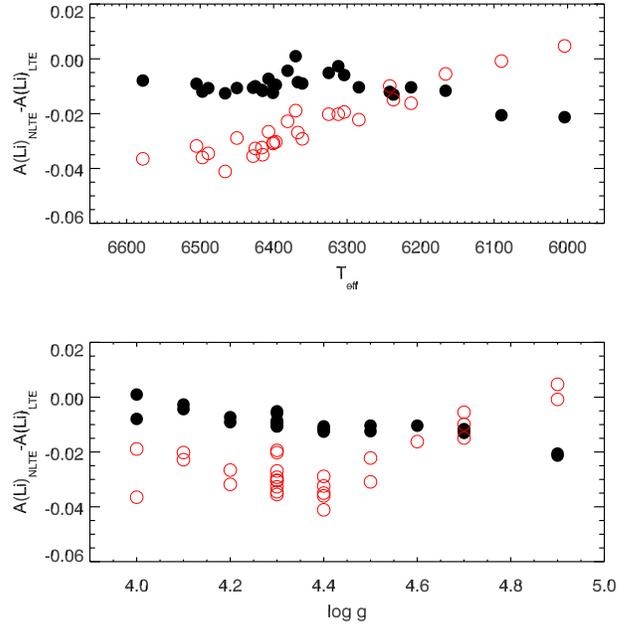}
      \caption{The NLTE correction (A(Li)$_{\mathrm 1D,NLTE}$ $-$ A(Li)
	      $_{\mathrm1D,LTE}$), computed for each star using our model atom
	      (black filled dots) along with the \citet{carlsson94} values (red
	      open circles), plotted against \teff\ and \glog. The 3D
	      temperature scale is assumed. } \label{compare_carlsson}
	      \end{figure}

A parallel grid of COG was produced using {\tt LHD} models sharing the same
parameters as the \cobold\ ones. The 1D syntheses were produced both
including and neglecting NLTE effects, for the specific purpose of deriving a
grid of NLTE corrections applicable to our 1D Li abundances. For comparison, 
Fig. \ref{compare_carlsson} shows our 1D NLTE corrections (for the 3D
temperature scale) versus \teff\ and \glog, together with the corresponding
values obtained by using the \citet{carlsson94} NLTE corrections, {while 
Fig. \ref{compare_lind} shows a similar comparison using the updated calculations by \citet{lind09}.
Since \citet{lind09} corrections are defined down to [Fe/H]=-3, in Fig. \ref{compare_lind} [Fe/H]=-3
is assumed for all stars both in computing our NLTE correction and in computing those based on
\citet{lind09} scale. Trends with effective temperature and gravity are extremely similar for our
corrections and those of \citet{lind09}, but a very uniform offset of about 0.03 dex is present 
between the two set of corrections. The origin of this offset is probably the 
different sets of underlying atmosphere models. Since the offset is quite
uniform across the sample, using either set of corrections is of no consequence on the scientific
output of the present work. }

\section{Results}

We decided to adopt the 3D temperature scale (and its derived parameters) together
with the 3D NLTE Li abundance set as our preferred values, and henceforth, when
not otherwise specified, we will refer to these. 

   \begin{figure}
   \centering
   \includegraphics[width=9cm]{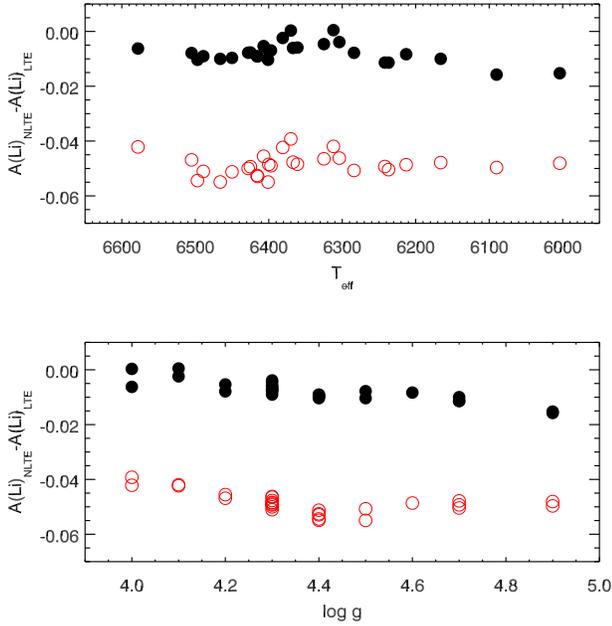}
      \caption{The NLTE correction (A(Li)$_{\mathrm 1D,NLTE}$ $-$ A(Li)
	      $_{\mathrm1D,LTE}$), computed for each star using our model atom
	      (black filled dots) along with the \citet{lind09} values (red
	      open circles), plotted against \teff\ and \glog. The 3D
	      temperature scale is assumed, and [Fe/H]=-3 is imposed for all stars.} \label{compare_lind}
	      \end{figure}

\subsection{Sensitivity to the adopted \teff\ scale}

One of the most remarkable results of this work is that, although the
choice of temperature scale alters the parameters and derived Li abundance of the stars,
it does not change the general picture that emerges. Table \ref{fitstable}
provides the results of Kendall's $\tau$-test and the slopes of linear fits to
the A(Li)-[Fe/H] and A(Li)-\teff\ relations. The linear fits were obtained taking
into account errors in both variables using the {\tt fitexy} routine
\citep{press}. The A(Li) error was assumed to be fixed at 0.09 dex, the error in
[Fe/H] to be given by the \ion{Fe}{i} line-to-line scatter for each star, and the error in
\teff\ to have a constant value of 130~K. The sample consists of 27 of the 28 stars
for which we have Li measurements, excluding 
\object{CS 22882--027}, for which we only have an upper 
limit to A(Li), as well as \object{CS 21188--033}, where the residuals from the best-fit 
regressions are on the order of 3-4$\sigma$.

In statistical terms, for all four temperature scales a non-parametric
Kendall's $\tau$-test indicates that A(Li) correlates with [Fe/H] at a very high
level of significance (see Table \ref{fitstable}). Moreover, a linear fit to the
A(Li) - [Fe/H] relation on the different temperature scales produces slope values
that are both always significant at the level of 3$\sigma$ and, strikingly, consistent with each
other within 1$\sigma$. The slope values are also the highest reported to date. The hypothesis that the ``slope'' in the
A(Li) - [Fe/H] relation might be due to the specific \teff\ scale chosen can
thus be safely rejected. 

\begin{figure}
\centering
\includegraphics[width=9cm]{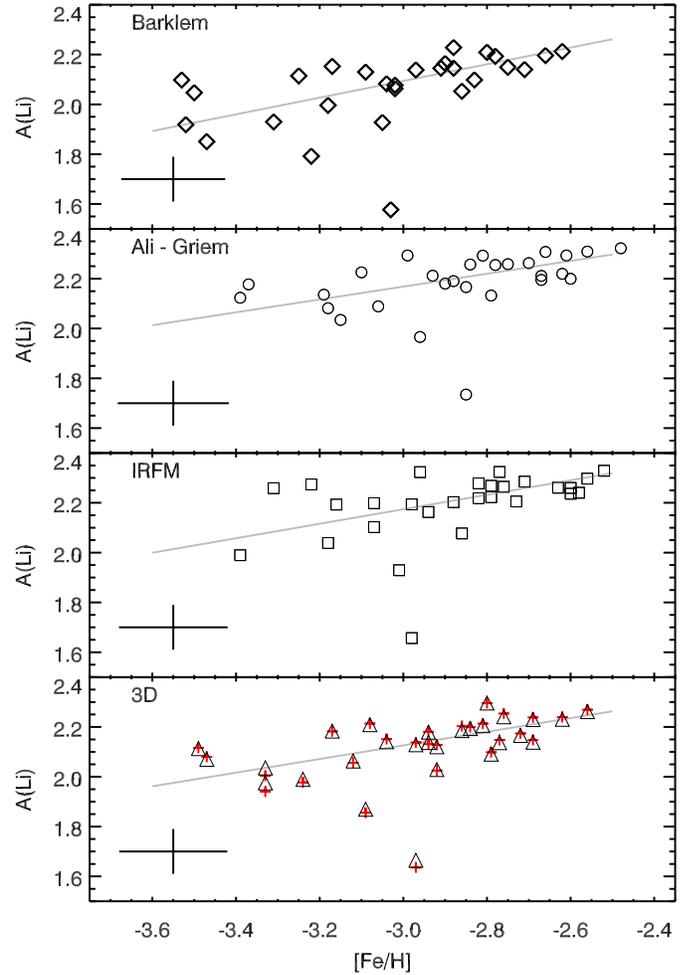}
\caption{Li abundance versus [Fe/H] for the four temperature estimates. 
Top to bottom, BA, ALI, IRFM, and 3D temperatures. For the 3D temperature scale,
the black triangles represent the 3D NLTE Li abundances, while the red crosses 
represent 1D LTE Li abundances with the NLTE corrections applied. 
The best-fit linear relation (as per Table  \ref{fitstable}) is indicated by a gray line. A typical error bar of $\pm$0.09 dex
in Li abundance and the average [Fe/H] error bar are also displayed. \label{livsfeh}}
\end{figure}

\subsection{The meltdown of the Spite plateau: slope or scatter?}

\label{slope_scatter}
Three different \halpha-based \teff\ scales, as well as the totally independent
 IRFM scale, concur in indicating that the Spite plateau is disrupted 
 below [Fe/H] $\sim -$3 (see Fig. \ref{livsfeh}). Close to that metallicity, one observes a significant increase
 in the Li abundance scatter, which appears to act {\em always towards lower
 abundances}. In other words, while some rare stars persist at the plateau level
 even at very low metallicity \citep[\object{CS 22876--032 A},][]{gonzalez08},
 the vast majority exhibit some degree of Li ``depletion'' (with respect to the
 plateau value). 
 
 \begin{table}
\caption{Parameters of the models in the 3D \cobold\ and 1D {\tt LHD} grids used
 in the 3D NLTE Li abundances, and in the computation of NLTE corrections. \label{3dgrid}}
\begin{center}
\begin{tabular}{lllcc}
\hline
{\bf \teff} & {\bf \glog} & {\bf [Fe/H]} & {\bf A(Li)$^a$} & {\bf EW$^b$}\\
{\bf K}     & {\bf cgs}   &              &           & {\bf pm} \\
\hline
5472. & 4.00 & $-$2.0 & 0.90 $-$ 2.10 &  5.71 $-$  69.71 \\
5479. & 4.50 & $-$2.0 & 0.90 $-$ 2.10 &  5.94 $-$  72.80 \\
5505. & 3.50 & $-$2.0 & 0.90 $-$ 2.10 &  4.98 $-$  61.94 \\
5846. & 4.00 & $-$3.0 & 1.30 $-$ 2.50 &  6.67 $-$  76.04 \\
5856. & 4.00 & $-$2.0 & 1.30 $-$ 2.50 &  6.79 $-$  77.83 \\
5861. & 3.50 & $-$2.0 & 1.30 $-$ 2.50 &  6.28 $-$  72.67 \\
5923. & 4.50 & $-$2.0 & 1.30 $-$ 2.50 &  6.17 $-$  73.16 \\
5924. & 4.50 & $-$3.0 & 1.30 $-$ 2.50 &  5.76 $-$  68.94 \\
6269. & 4.00 & $-$3.0 & 1.30 $-$ 2.50 &  3.29 $-$  43.33 \\
6272. & 4.50 & $-$3.0 & 1.30 $-$ 2.50 &  3.32 $-$  43.83 \\
6278. & 4.00 & $-$2.0 & 1.70 $-$ 2.90 &  8.05 $-$  86.35 \\
6287. & 3.50 & $-$2.0 & 1.70 $-$ 2.90 &  7.42 $-$  81.33 \\
6323. & 4.50 & $-$2.0 & 1.70 $-$ 2.90 &  7.71 $-$  84.35 \\
6408. & 4.00 & $-$3.0 & 1.30 $-$ 2.50 &  2.68 $-$  36.35 \\
6533. & 4.50 & $-$2.0 & 2.10 $-$ 3.30 & 13.55 $-$ 119.59 \\
6534. & 4.00 & $-$2.0 & 2.10 $-$ 3.30 & 12.87 $-$ 118.58 \\
6556. & 4.50 & $-$3.0 & 1.30 $-$ 2.50 &  2.16 $-$  30.20 \\
\hline
\end{tabular}
\end{center}
{\scriptsize $^a$ Minimum and maximum A(Li) covered in the COG \newline
$^b$ Minimum and maximum EW for the Li doublet in the COG}
\end{table}

Two stars in the sample exhibit anomalously low Li abundances. The star \object{CS 22188--033}
exhibits a mild Li depletion with a 3D NLTE A(Li) = 1.66, while
\object{CS 22882--027} has no detectable Li doublet (see sect. \ref{rogue_stars}).

One of the much-debated results concerning the behavior of Li abundances in
metal-poor halo dwarfs has been the reported existence of a correlation between A(Li)
and [Fe/H], since it was first reported by \citet{ryan99}. We investigated this 
by means of two different statistical tests. The Kendall's $\tau$
rank-correlation test attempts to detect a (positive or negative)
correlation, and has the fundamental strength of being non-parametric. In other
words, it does not attempt to look for a specific relation to fit the data.
As seen above, Kendall's $\tau$-test quite strongly supports
the existence of a correlation. 

The other obvious strategy we adopt is to fit the data with a linear function
and see whether the slope found is statistically significant. This significance
might be weakened if the data are indeed correlated, but the underlying relation
is {\em not} linear. In our case, again, the slope of the linear relation is
significant at $3\sigma$ for all the considered temperature scales. This
a finding, however, does {\em not} imply that the underlying
``physical'' relation between [Fe/H] and A(Li) is linear, as it would be, for
example, if there was a constant Li production with increasing [Fe/H].

\begin{table}
\caption{Kendall rank correlation probability, intercepts, and slopes of 
the linear fit and standard deviation of the slope for the plots shown in Figs. \ref{livsteff} 
and \ref{livsfeh}. The star \object{CS 22188--0033} has been excluded from the
fit.}

\label{fitstable}      % is used to refer this table in the text
\centering                          % used for centering table

\begin{tabular}{l c c c }        % centered columns (4 columns)
\hline
{\bf Parameter} & {\bf correlation} & {\bf linear fit} & {\bf linear fit} \\
{\bf set }      & {\bf probability} & {\bf intercept } & {\bf slope     } \\
\hline
A(Li) vs. \teff & & &\\
\hline                                   %inserts single line
BA              & 0.978             & $-$3.038$\pm$1.542 & 8.17e$-$4$\pm$2.33e$-$4  \\
ALI             & 0.962             & $-$7.917$\pm$10.83 & 1.57e$-$3$\pm$1.00e$-$3  \\
IRFM            & 0.993             & $-$2.133$\pm$1.448 & 6.68e$-$4$\pm$2.13e$-$4  \\
3D              & 0.990             & $-$2.910$\pm$1.898 & 7.92e$-$4$\pm$2.78e$-$4  \\
\hline
\hline
A(Li) vs. [Fe/H] & & &\\
\hline
BA              & 1.000             & 3.099$\pm$0.245 & 0.335$\pm$0.080  \\
ALI             & 0.999             & 2.942$\pm$0.245 & 0.258$\pm$0.085  \\
IRFM            & 0.998             & 3.047$\pm$0.268 & 0.291$\pm$0.092  \\
3D              & 0.999             & 2.948$\pm$0.248 & 0.274$\pm$0.083  \\
\hline
\end{tabular}     
\end{table}

To shed more light on the issue, in Fig. \ref{residuals} we plot
the residuals of the best-fit A(Li) versus [Fe/H] relation, as listed in Table
\ref{fitstable}. An increase in the scatter below [Fe/H]$\sim$ $-$2.8 was
already visually apparent in Fig. \ref{livsfeh}, and remains clearly recognizable
in Fig. \ref{residuals} once the best-fit linear relation is subtracted. To
provide quantitative estimates of the level of scatter, we divided the sample into
two in terms of metallicity, a metal-richer subsample including the 8 stars with
[Fe/H]$_{\rm 3D}>-2.8$, and a metal-poorer sub-sample including the 19 stars 
below that threshold. The star \object{CS 22188--033} is plotted
in the figure, but it has not been considered in this computation. We then
computed the dispersion in the residuals of the two subsamples for the four
temperature scales: $\sigma_{\rm hi, BA}=0.04$ dex, $\sigma_{\rm lo,BA}=0.10$
dex; $\sigma_{\rm hi,ALI}=0.05$ dex, $\sigma_{\rm lo,ALI}=0.08$ dex;
$\sigma_{\rm hi, IRFM}=0.02$ dex, $\sigma_{\rm lo,IRFM}=0.10$ dex; 
$\sigma_{\rm hi,3D}=0.05$ dex, $\sigma_{\rm lo,3D}=0.09$ dex. For every
temperature scale, the scatter in the residuals is about twice as large (or
more) below [Fe/H]=$-$2.8 than above. It is thus clear that the tight, flat
relation that is known as the Spite plateau develops both a tilt and a significant 
scatter at low metallicities. Once again, it is remarkable how
the level of scatter appears independent of the assumed temperature scale. 

\begin{figure}
\centering
\includegraphics[width=9cm]{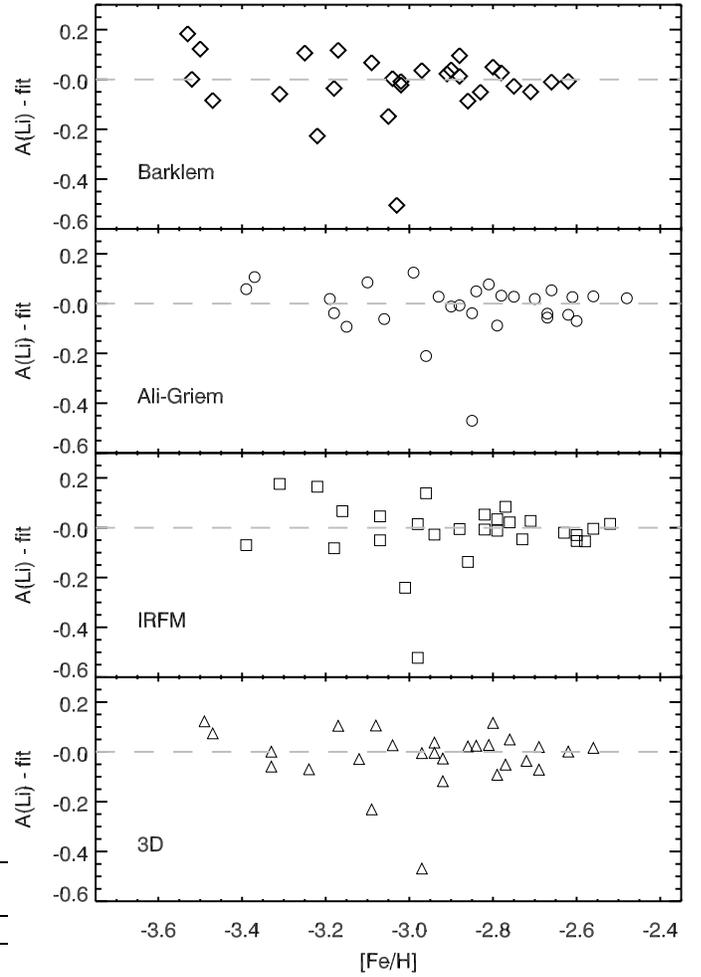}
\caption{Residuals of the best fit of A(Li) vs. [Fe/H] listed in Table \ref{fitstable}.
}
\label{residuals}
\end{figure}

We note that the true Li doublet EW does not change much with
metallicity, since in general A(Li) does not vary by more than 0.2 dex. The
quality of the Li doublet measurement is thus roughly constant across the whole
metallicity range. The increase in scatter thus cannot be attributed to the
declining quality of the measurements. On the other hand, \ion{Fe}{i} and
\ion{Fe}{ii} lines do become weaker with metallicity, which lowers the quality
of the gravity estimation. Since the \halpha\ line is quite gravity sensitive,
inaccurate gravities reflect directly on \halpha-based \teff\ estimations, and
thus on A(Li). On the other hand, the IRFM temperature scale is totally
insensitive to this effect, and yet shows the largest increase in the scatter of
its residuals, and a low-metallicity scatter equal to those of the
\halpha-based \teff\ scales. This reinforces our impression that the increase in the
A(Li) residual scatter should indeed be real.

\subsection{Plateau placement}

\label{plateau_placement}
As a consequence of what is said above, it hardly makes sense to provide an
average value for A(Li) in our stars. One might still try, however, to determine
the position of the plateau for the more metal-rich stars of the sample, which
still appear to fall onto it. Every operation of this kind is somewhat arbitrary, since
there is no clear-cut transition between the plateau at higher Fe content and
the sloping / dispersed distribution at low metallicity. We thus decided to
employ the 9 stars whose metallicity is equal or greater than $-$2.8 {\em in the
3D scale}, and consider their A(Li) and dispersion as being representative of the
Spite plateau. The resulting values are $\rm\left\langle A(Li)
\right\rangle=2.164\pm0.059$ in the BA scale, $\rm\left\langle A(Li)
\right\rangle=2.261\pm0.053$ in the ALI scale, $\rm\left\langle A(Li)
\right\rangle=2.264\pm0.044$ in the IRFM scale, and $\rm\left\langle A(Li)
\right\rangle=2.199\pm0.086$ using A(Li)$_{\rm 3D,NLTE}$ together with the 3D
temperature scale.

\subsection{The A(Li) -- \teff\ correlation}
\label{li_teff_corr}

In Fig. \ref{livsteff}, as well as in Table \ref{fitstable}, a positive slope of Li
abundance with effective temperature is present in all the temperature scales. It is, however, remarkable that this slope
is driven by the coolest stars. Adopting our 3D temperature scale, if we select
only the stars hotter than 6250~K (22 stars), the slope essentially vanishes, the
probability indicated by Kendall's $\tau$ drops to 89\%, and a parametric test
does not detect any slope.

\begin{figure}
\centering
\includegraphics[width=9cm]{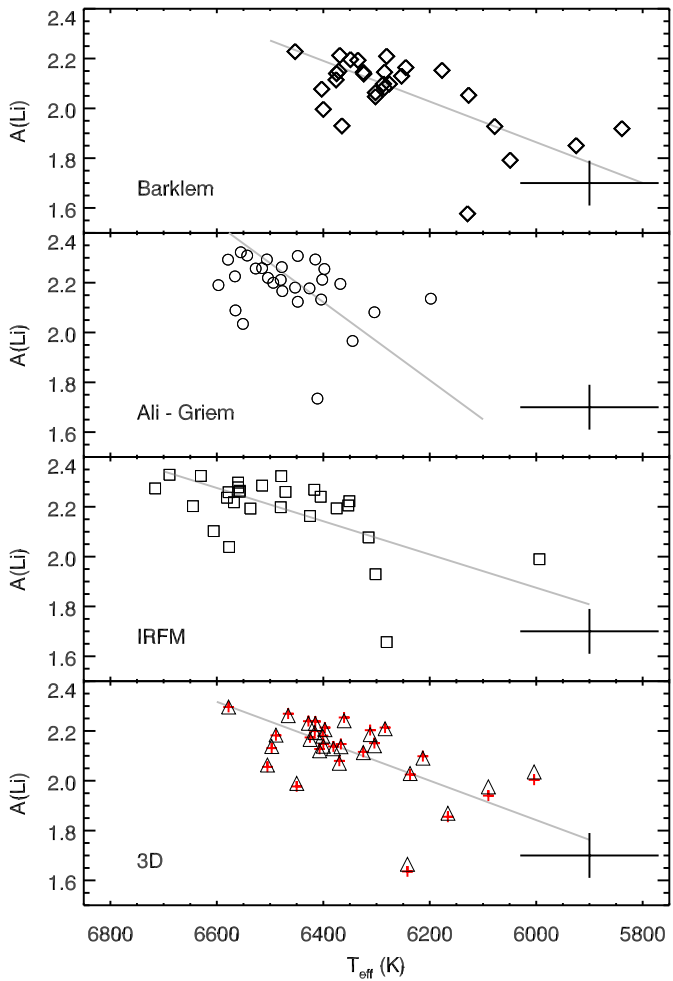}
\caption{Li abundance versus effective temperature for the four temperature 
estimates. Symbols are the same as in Fig. \ref{livsfeh}. Typical error bars of
$\pm$0.09 dex in the Li abundance and $\pm$130 K in \teff\ are also displayed. The
best-fit linear relation as per Table \ref{fitstable} is indicated by a gray
line. }
\label{livsteff}
\end{figure}

Given the very small population of this temperature range in our sample, it is possible that we just
missed any undepleted cool object. On the other hand, this might suggest that the
decline (caused by convection) usually seen for stars cooler than $\sim
5700$\,K at higher metallicities may set in at higher temperatures for EMP
stars \citep[although observationally the opposite seems to be true, see][]{boesgaard05}. 

We emphasize, however, that the depletion of the cool end of the sample  {\em is not driving} the A(Li)-[Fe/H] correlation: removing the aforementioned five cool
stars has a negligible effect on this result. On the 3D scale, the Kendall's $\tau$
correlation probability of A(Li)$_{\mathrm{3D,NLTE}}$ with [Fe/H] passes from 0.999 to 0.998, while
the slope of the linear fit goes from 0.274$\pm$0.083 to 0.253$\pm$0.086 when these 5 stars are removed. This is because
the cool stars, while appearing to be all Li depleted, are evenly distributed in metallicity between [Fe/H]$\sim$-3.2 and -2.8. On the other hand, 
removing these stars would affect the detected increase in the scatter at lower metallicities: applying the same residual analysis mentioned in Sect. \ref{slope_scatter}, but now
on the hot stars only, would yield $\sigma_{\rm hi,3D,hot}=0.04$ dex (was 0.05 with the full sample), and $\sigma_{\rm lo,3D,hot}=0.06$ dex (was 0.09 with the full sample).

\subsection{\object{CS 22882--027} and \object{HE 1148--0037}}
\label{rogue_stars}

Two additional stars were included in the original sample, but Li abundance has
not been computed for them, for different reasons.  We elaborate briefly on
these objects. 

\begin{figure}
\centering
\includegraphics[width=9cm]{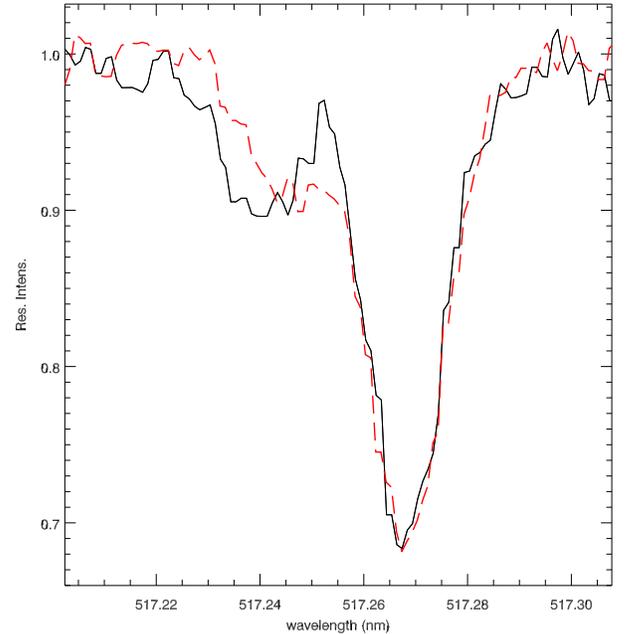}
\caption{The \ion{Mg}{i} 517.268 nm line for \object{HE 1148--0037}, 
in spectra taken at JD 2453787.6803 (black continuous line) and JD 2453823.5443
(red dashed line), separated by 35.86 days. The spectra have been normalized,
and Doppler shifted to bring the primary component to a rest-frame wavelength. The
binarity is readily visible, as well as the variation in the components'
separation. \label{he1148}}
\end{figure}

Atmospheric parameters and metallicity were computed for \object{CS 22882--027}
(see Table \ref{params}), but the star has no detectable Li doublet (see Fig.
\ref{spectrasamples}). Given a S/N$\sim$80, as measured in the Li doublet range,
a typical line FWHM of 0.033 nm, and a pixel size of 0.0027 nm, the \citet{cayrel88}
formula predicts that a Li doublet of EW=0.563 pm would be measured at
3$\sigma$ confidence (1$\sigma$=0.187 pm). Employing our fuctional fit, this leads
to an upper limit of A(Li)$_{\rm 3D,NLTE}\le$ 1.82, assuming an EW=0.563 pm and A(Li)
$_{\rm 3D,NLTE}\le$1.34, assuming EW=0.187 pm. We have a single-epoch spectrum
for this star that shows no sign of a double-line system.

The star \object{HE 1148--0037} was also originally included in the sample, 
but immediately set aside, since from visual inspection of the two available
spectra it turned out to be a double-lined binary system. As shown in Fig.
\ref{he1148}, the two spectra, separated by 35.86 days, clearly exhibit 
evidence of the double-line system, as well as readily recognizable variation in the separation between
the two line systems. We were able to retrieve 3 spectra of \object{HE
1148--0037}, and measure radial velocities for the two components, which
are given in Table \ref{vrad_he1148}, by means
of cross-correlation against a synthetic template (\teff=6000 K, \glog=4.0,
[Fe/H]=$-$2.0). In our two spectra (2006-02-21 and 2006-03-29), cross-correlation 
was computed in the 490nm-570nm range. We also had the lower
resolution, blue-range only HERES spectrum, which was used to derive the
V$_{\rm rad}$ after masking all the broad hydrogen lines. Internal errors of the
radial velocity estimate were evaluated by performing a Monte Carlo test on a sample of 50
simulated binary star spectra with noise added to emulate a S/N=90,
the component separation and resolution being equivalent to that of the March 29, 2006
observation. The test inferred an average error of 0.072 km/s for the primary
component and 0.246 km/s for the secondary. These values are
representative of the internal errors in the cross-correlation procedure, but
are surely dominated by the spectrograph zero-point calibration uncertainty,
which was not taken into account particularly well, because very precise radial velocities were
not among the goals of this study. We thus list in Table \ref{vrad_he1148} an
estimated total error of 1 km/s for all our observations, which we take to be
representative of the overall systematics of these measurements.

On the other hand, \citet{aoki09} report AN uncertainty given by the internal scatter in
the radial velocities obtained from different lines; they do not take into account the
systematic uncertainties in the wavelength calibration, hence the much lower value of uncertainty. In
Tab. \ref{vrad_he1148}, we kept the value they provide, but we propose that the true uncertainty 
is again close to 1 km/s. If we consider their measure to be 
representative of the primary component radial velocity, it reproduces well the three other
measurements we present.

\begin{table}
\caption{Barycentric radial velocities for the two components of the binary system \object{HE1148--0037} as measured from the spectra available to us.
\label{vrad_he1148}}
{\small
\centering
\begin{tabular}{lllll}
\hline
{\bf date} & {\bf hour} & {\bf JD}         & {\bf V{\boldmath $_{rad}$}} & {\bf V{\boldmath $_{rad}$}}  \\
{\bf UT}   & {\bf UT}   & {\bf UT}         & {\bf km/s}                  & {\bf km/s}                   \\
           &            &                  & {\bf primary}               & {\bf secondary}              \\
\hline
8 May 2003$^a$ & 11:43:06   & 2452767.9882 & $-$36.16$\pm$1                & $-$3.35$\pm$1                  \\
27 Feb. 2005$^b$ & 10:33:36   & 2453428.94   & \multicolumn{2}{c}{$-$10.88$\pm$0.16}                        \\
21 Feb. 2006& 04:19:40   & 2453787.6803     & $-$9.15$\pm$1                 & $-$22.37$\pm$1                 \\
29 Mar. 2006& 01:03:46   & 2453823.5443     & $-$6.90$\pm$1                 & $-$22.56$\pm$1                 \\
\hline
\multicolumn{5}{l}{\scriptsize $a$ HERES spectrum}\\
\multicolumn{5}{l}{\scriptsize $b$ \citet{aoki09} single radial velocity}\\
\end{tabular}}
\end{table}

\subsection{Comparison with other results}
\label{compare_others}

Our most significant overlap is of course with the
\citet{bonifacio07} sample, of which the present study represents a continuation. Of the 19
stars in \citet{bonifacio07}, 17 were reanalyzed here (the two remaining
stars, \object{BS 16076--006} and \object{BS 16968--061} turned out to be
subgiants and have thus been dropped). \citet{bonifacio07} determined \teff\ by fitting the
\halpha wings fits with profiles synthesized using the \citet{barklem00,
barklem00b} self-broadening theory. Their temperature scale thus closely
resembles our BA scale, but the \halpha\ gravity sensitivity was not taken into
account in that study -- \glog=4.0 was assumed in computing the profiles. The
effect is clearly visible in Fig. \ref{temp_differ}.  The upper panel of this
figure plots the temperature difference,\teff (this work)$-$\teff (\citealt{bonifacio07}),
against the \citet{bonifacio07} gravity estimate. It is immediately evident
that, for stars around \glog=4.0, the temperature difference approaches zero.
For stars at higher gravities, our \teff\ estimate is below the
\citet{bonifacio07} value by up to 200~K. This reflects the behavior presented in
Fig.\ref{thadif}: when a profile is fitted with a synthetic grid computed
for a gravity that is {\em underestimated}, this leads to an {\em
overestimated} \teff. We note that Fig. \ref{temp_differ} does not
tell the entire story. As we do here, \citet{bonifacio07} estimated \glog by enforcing Fe ionization equilibrium, 
so the derived gravity values are not identical in this
work and \citet{bonifacio07}. High-gravity and low-gravity stars do, however, 
retain their approximate placement in both cases, although the gravity span can
be somewhat stretched by the \teff\ bias. The effect of the \teff\ difference on
[Fe/H] is shown in the middle panel of Fig. \ref{temp_differ}, while in the
lower panel the difference between A(Li) for the same stars is plotted,
considering here the LTE values (to eliminate the effect of the marginally
different NLTE corrections applied in the two works). The results shown are to be
expected, given the strong \teff\ sensitivity of the \ion{Li}{i} 670.8 nm
doublet: the current A(Li) is higher by up to about 0.1 dex for low-gravity
stars, while it is lower by roughly the same amount for high-gravity stars. On
the other hand, it is easy to see how the discrepancy will only marginally
affect a linear fit of A(Li) versus [Fe/H]: the stars are displaced roughly along a
1:1 diagonal in the A(Li) -- [Fe/H] plane.

\begin{figure}
\centering
\includegraphics[width=9cm]{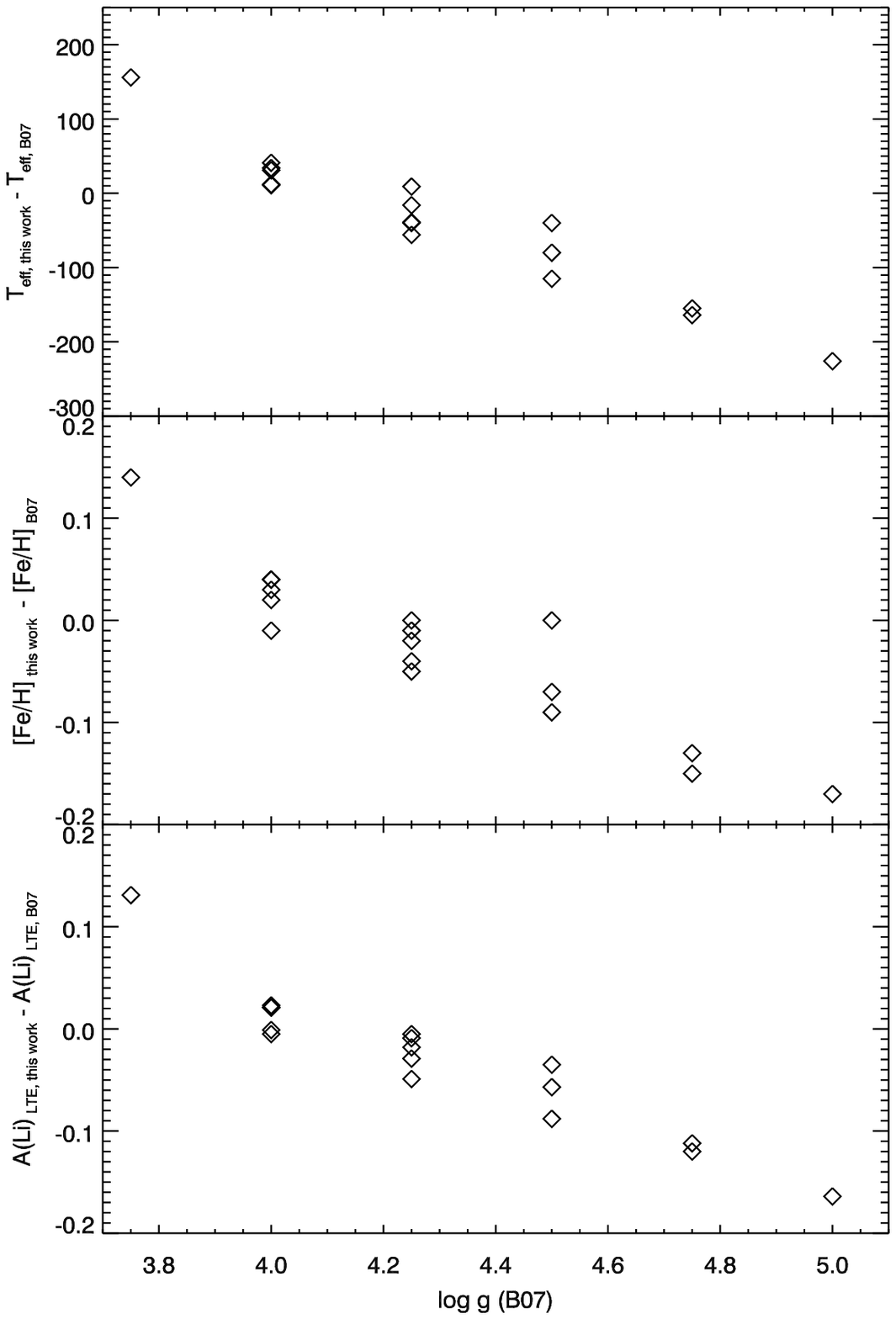}
\caption{Comparison between the results of this work and of \citet{bonifacio07} for the 17 stars in common. In the upper panel, the difference between our \teff\  and the \citet{bonifacio07} \teff\ determination is plotted against the value of \glog\ in \citet{bonifacio07}. In the center panel, we show [Fe/H] difference, and in the lower panel, the difference in the A(Li) LTE. For our results, BA temperature scale is used.
}
\label{temp_differ}
\end{figure}

The star \object{LP 815--43} is the only object that overlaps with the
\citet{asplund06} sample. In that work, effective temperature is measured again
by fitting \halpha\ wings with a set of synthetic profiles. The details of
the fitting procedure differ somewhat and the synthetic profiles are computed
from {\tt MARCS} models using the {\tt BSYN} synthesis code. The \citet{barklem00,
barklem00b} self-broadening theory is assumed here for \halpha, so again the BA scale
is the one to be used in the comparison. The \halpha\ gravity
sensitivity is taken into account here, and the derived \teff\ is quite close to
our value (6400 K vs. 6453 K in this work). \citet{asplund06} determine
metallicity from \ion{Fe}{ii} lines, and gravity from Hipparcos parallaxes, but
again the values do not differ much from our results (\glog=4.17, V$_{\rm
turb}$=1.5, [Fe/H]=$-$2.74). The residual 0.14 dex offset in metallicity is in
good agreement with the 0.2 dex offset detected by \citet{bonifacio07} between
their metallicity scale and that of \citet{asplund06}. The lithium abundance
is again quite close to our result: from the 670.8 nm doublet, they derive
A($^7$Li)=2.16, while our value is 2.23. The difference is fully accounted for
once the \teff\ effect is considered (0.03 dex) as well as the already known,
albeit still unexplained, 0.04 dex bias between A(Li) as derived by means of {\tt
BSYN} and {\tt turbospectrum} \citep{bonifacio07}.

The same star is also in common with \citet{hosford09}. That work uses
excitation equilibrium to estimate \teff, a method that is not directly
comparable with any of our temperature scale. 
%The parameters they derive (\teff=6529, \glog=4.40 V$_{\rm turb}$=1.4, [Fe/H]=-2.61 for the MS scale, \teff=6400K) are fairly close to the ones of our 3D and ALI temperature scales, but with a appreciably higher gravity and a [Fe/H] higher by 0.19 dex. 
The authors derive two parameter sets, one assuming the star belongs on the main
sequence, and another assuming it is a subgiant. In the first case (\teff=6529,
\glog=4.40, V$_{\rm turb}$=1.4, [Fe/H]=$-$2.61), they obtain a temperature 
that is only about 50~K cooler than for our 3D and ALI scale, but a higher gravity and metallicity. In
the second case (\teff=6400, \glog=3.80, V$_{\rm turb}$=1.4, [Fe/H]=$-$2.68),
they derive a temperature 50~K cooler than for our BA scale, the same gravity,
but again to a metallicity 0.2 dex higher. Their measured Li doublet equivalent
width is about 0.2 pm smaller than that we find, and in both cases they derive
an A(Li) that is about 0.1 dex smaller than ours. This is consistent with
the expected combined effect of the difference in both \teff\ and the Li doublet
EW.

Three stars are in common with the \citet{aoki09} sample, but we determined
A(Li) for only two of them, \object{CS 22948--093} and \object{CS 22965--054}.
\citet{aoki09} employ \halpha- as well as H$\beta$-wing fitting to determine
\teff\, on the basis of {\tt MARCS} models and using the \citet{barklem00,
barklem00b} self-broadening for \halpha. Once more, our BA temperature scale is
the one most appropriate for a comparison. The \halpha\ gravity sensitivity is taken into account.
Surface gravities are estimated by comparing with isochrones as well as by
evaluating the \ion{Fe}{i}--\ion{Fe}{ii} ionization equilibrium. The
parameter values they derive for \object{CS 22948--093} are in close agreement with those
we derive (\teff(\halpha)=6320 K, \teff=6380, \glog=4.4, V$_{\rm turb}$=1.5),
while [Fe/H] is 0.13 dex lower at $-$3.43. The derived value of A(Li)$_{\rm LTE}$=1.96 is inexcellent agreement with our
value of 1.935. \object{CS 22965--054} shows a more
significant discrepancy in \teff. Their value of \teff(\halpha)=6390K is remarkably
higher than our, while \teff(H$\beta$) is much closer to our value. Since the adopted temperature is the average of the two, the final \teff\ is ultimately just 56 K hotter than our value. Their derived
surface gravity is also
very close in value (\glog=3.9), while V$_{\rm turb}$ is the same. However, \citet{aoki09} derive
the same lithium abundance as we do, A(Li)$_{\rm LTE}$=2.16, due to their lower
measured value of EW (2.03 pm) for the Li doublet.

The third star in common with \citet{aoki09} is \object{HE 1148--0037}, for which
\citet{aoki09} do not appear to have noticed its binarity. On close inspection,
the single spectrum they employed (JD~2453428.94) shows signs of line asymmetry
(Aoki 2009, priv. comm.), but it appears to have been taken quite close to conjunction.
In contrast, as seen in Sect. \ref{rogue_stars}, the three spectra we have of
this star all show quite clearly the two line systems.     

{We observed No star in common with \citet{melendez04}. A comparison with their sample
is nevertheless interesting due to the extension of their sample to very low metallicities,
despite with a limited number of stars (a total of 10 stars were analyzed below [Fe/H]=-2.5,
4 at or below [Fe/H]=-3).
\citet{melendez04} detected no slope in the Spite plateau, for which they advocated a
high value of $\rm\left\langle A(Li)\right\rangle$=2.37. As noticed in \citet{gonzalez09}, 
the \teff\ scale adopted by
\citet{melendez04} is systematically hotter than the one we employ for metal poor dwarfs, 
on average by 87 K. This explains half of the descrepancy between our average IRFM plateau placement
and their own, and can account in principle for their failure to detect the slope, assuming their
temperature scale and ours diverge progressively at low metallicities. 

Lithium abundances for two extremely metal poor stars (\object{HE 0233--0343} and \object{HE 0945--1435}) 
were recently presented by \citet{garcia08}. Both stars show extremely low Fe content ([Fe/H]$\sim$-4),
but probably because of the weakness of \ion{Fe}{ii} lines, the estimation of gravity is uncertain. This
affects the determinations of both the evolutionary status (either MS or early SGB) and \teff, which is 
derived from \halpha\ wing fitting in a way similar to that used with our BA scale. The stars appear to 
be fairly cool, 6000K$\le$\teff$\le$6250K, which would place them among the ``cool stars'' of our sample 
as described in Sect. \ref{li_teff_corr}, and  both objects show significantly depleted Li, A(Li)$\sim$1.8.
Owing to the uncertainty of the parameters determination we did not include these stars in Fig. \ref{alltogethernow}.
}

\section{Possible biases}                                                                                     

\subsection{Binary stars}
Two potential biases can, in principle, be responsible for producing
systematically low Li abundances, and a trend of A(Li) with metallicity. The
first one is of course the presence of undetected binaries, for which the
veiling by the secondary star will systematically reduce the EW of the lines of
the primary, leading to an underestimate of both metallicity and Li abundance.
The true impact of this effect is difficult to evaluate, mainly because little
is known about the fraction, and mass-ratio distribution, of binaries at low
metallicities. \citet{duquennoy91} report, for G dwarfs in the solar vicinity, a
fraction of 44\% of stars having a companion with $q=M_2/M_1>0.1$, about 1/3 of
which have $q>0.5$. \citet[and references therein]{latham02} found
that the halo binary population does not differ significantly from the disk
one, although we note that they did not explore significant
numbers of stars with metallicities as low as the stars in our present sample.

It nevertheless seems unlikely that undetected binaries 
pollute our sample significantly. We checked for binarity by inspecting the Mg b triplet
lines (e.g., see Fig. \ref{he1148}). Our spectra have a typical S/N$\sim$100 or
higher, and Mg b lines have a typical EW of 10 pm. A line with central residual
intensity of 0.95 would be detected at least at the 5$\sigma$ level in this typical
spectrum, and have a typical EW of 0.8 pm. As per \citet{gonzalez08}, to reduce
a 10 pm line to 0.8 pm, a ratio of the continua fluxes of about 11.5 is needed.
If we roughly assume that the total luminosity scales accordingly, this
corresponds roughly to $q=0.5$ (since $L\propto M^{3.2}$ on the main sequence,
see \citealt{kippenhahn90}). A similar flux ratio in the Li doublet range would
lead to a correction of the Li doublet EW for the primary star of about 8\%,
corresponding to 0.03~dex in A(Li). In other words, every binary star requiring
significant veiling correction on the primary spectrum would also be promptly
detectable because of the double line system. This system could only stay
undetected if the radial velocity separation of the two stars was quite small at
the moment of the observation(s), so that the two line systems remained blended.

In addition to the above, if we take the figures of \citet{duquennoy91} at face value,
find that about 13.5\% of the binaries are characterized by a
significant veiling of the primary (i.e. $q>0.5$). Our original sample comprised
30 stars, which implies that there are 4 expected ``significant binaries''. Three
stars have already been rejected from the sample, one of them (\object{HE
1148--0037}) being indeed a binary. Two more (\object{CS 22882--027} and \object{CS
22188--033}) exhibit significant lithium depletion or no Li doublet at all, and
were excluded from all statistical analyses. They are clearly the most likely
candidates to be binaries, albeit neither one shows a double line system\footnote{We have another UVES spectrum of \object{CS
22188--033}, taken at a different epoch, which does not show radial velocity variations with respect to the one used in the present work, nor signs of a double line system.}. One
could thus expect one more ``disguised'' binary to be biasing the sample. While
this is quite possible, it would hardly influence any of our results.

\subsection {3D NLTE effects on Fe ionization equilibrium}

The second problem relates to the use of \ion{Fe}{i}-\ion{Fe}{ii} ionization
equilibrium to estimate gravity. From preliminary computations, it appears that
3D corrections of Fe lines with excitation potentials of the same order as
employed in the present work could be quite large at low metallicities for stars
similar to those that we study. Moreover, corrections for \ion{Fe}{i} appear to be
negative (of about 0.2 dex), while they are positive (about 0.1 dex) for
\ion{Fe}{ii} lines, for a \teff=6500, \glog=4.5, [Fe/H]=$-$3.0 star. The
phenomenon is mainly caused by the overcooling that 3D treatment produces in the
outer layers of atmospheres at low metallicities, and appears to be of similar 
magnitude at [Fe/H]=$-$2 (due to the stronger saturation of Fe lines, which
drives their contribution function to higher layers), but would most likely
disappear above. If taken at face value, a 0.3 dex \ion{Fe}{i}-\ion{Fe}{ii}
imbalance would lead to an {\em overestimate} of \glog\ of about 0.5 dex  
when analyzed using 1D LTE models (as is our case regarding
metallicity and gravity estimation). We do indeed find higher gravities than 
expected from evolutionary tracks. On the other hand, we do not
currently have a 3D NLTE spectrosynthesis code for iron; we are thus unable to
account for NLTE effects, which are likely to counterbalance the 3D effect because 
of over-ionization occurring in the upper layers where overcooling is present in 3D
models. A similar mechanism is indeed active for Li, whose 3D LTE
abundance derived by \cobold-{\tt LINFOR3D} is about 0.2 dex below the
corresponding 1D NLTE value, while the 3D NLTE one is essentially
indistinguishable from the 1D NLTE result (Figs. \ref{livsfeh} and \ref
{livsteff}). If some degree of imbalance remains after NLTE is taken into
account, and is metallicity sensitive, this might introduce a bias when
\teff\ is determined by \halpha-wing fitting. It is indeed intriguing to note how
lower gravity estimates lead to higher temperatures, and as
a consequence, (somewhat) higher the Li abundances. On the other hand, the
IRFM temperature scale should be immune to this
problem,  A(Li) being quite insensitive to \glog\ itself, and the IRFM-based
analysis inferring abundances similar to the \halpha-based estimate. 

\section{Conclusions}

We have presented the largest sample to date of Li abundances for EMP halo
dwarf stars (27 abundances and one upper limit), including the largest sample to
date below [Fe/H] = $-$3 (10 abundances). Lithium abundance determination is
highly sensitive to biases in the effective temperature scale, and we have tried to
account for this using four different temperature estimators. In an additional
effort to accurately represent the stellar atmospheres of the sample stars, 3D,
time-dependent, hydrodynamical atmosphere models have been used to determinine
our preferred \halpha-based temperature scale, and a detailed 3D NLTE
spectrosynthesis has been applied to the determination of lithium abundance.
Both these techniques have been employed here for the first time, to our knowledge, in
the analysis of EMP stars. This has also allowed us to develop a useful fitting
formula allowing one to derive A(Li)$_{\rm 3D, NLTE}$ directly as a function of EW,
\teff, \glog, and [Fe/H] for EMP turn-off and early subgiant stars (see Appendix
\ref{fitappendix}). 

The first obvious conclusion of this work is that we have confirmed what was merely
suggested by the analysis of \citet{bonifacio07}, and previous works, that
at the lowest metallicity there is sizable dispersion in the Li abundances and
that there is a trend of decreasing Li abundance with decreasing metallicity. We
have also shown that these two conclusions do not depend on the
adopted temperature scale, as suggested by \citet{molaro08}. The results hold,
qualitatively, using both IRFM temperatures and H$\alpha$ temperatures,
regardless of the broadening theory adopted and irrespective of the use of either 1D or 3D
model atmospheres. Quantitatively, the results differ in the mean level of the Li
abundance, while the slopes in the A(Li) versus [Fe/H] relations agree within errors. None of the
temperature scales investigated produces a ``flat'' Spite plateau over the
full range in [Fe/H] (see Table \ref{fitstable}).

Our results are in substantial agreement with those of \citet{aoki09}. While
these authors do not detect a slope with either effective temperature or
metallicity, this happens simply because of the small extent of their sample in
both these parameters. On the other hand, they do point out that their sample
has a lower Li abundance than that observed at higher metallicities.

\begin{figure}
\centering
\includegraphics[width=9cm]{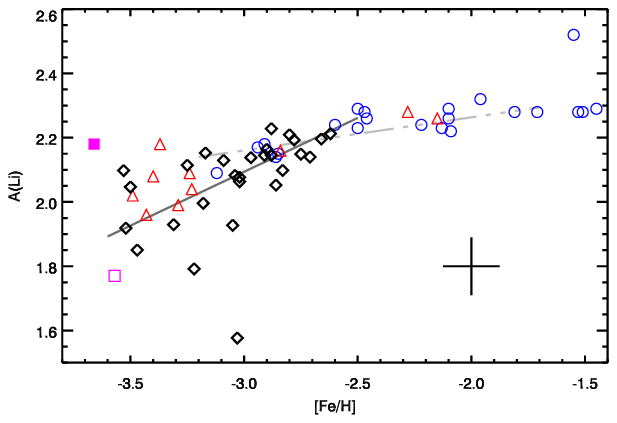}
\caption{A unified view of A(Li) vs. [Fe/H] from some studies for which a common temperature scale can be assumed. Blue circles, \citet{asplund06} data, red triangles, \citet{aoki09} data, magenta squares, \object{CS  22876--032} from \citet{gonzalez08}, filled symbol primary star, open symbol secondary star. Black diamonds, this work, BA temperature scale. Dot-dashed gray line, best linear fit to \citet{asplund06} data, continuous dark gray line, best fit to our data. Typical error bars for our data are displayed.}
\label{alltogethernow}
\end{figure}

The picture outlined by the aforementioned results acquires more significance, once we place it in
a broader context among the latest studies regarding lithium in EMP stars. Figure
\ref{alltogethernow} compares our results with those of three 
investigations employing compatible temperature scales. In this figure, open
blue circles represent stars from the \citet{asplund06} sample, red triangles
from the \citet{aoki09} data, and the two magenta squares the two components of
the double-lined binary system \object{CS 22876--032} \citep[][the filled square
corresponds to the primary star]{gonzalez08}. Our data are represented as black
diamonds (the results of the BA scale are shown, for compatibility with the
temperature scales used in the other three works)\footnote{\citet{gonzalez08}
derived \teff\ from photometry and isochrones, but a cross-check with \halpha\
profiles computed in 1D with \citet{barklem00} broadening confirmed the
result.}. The best linear fit to our data is shown as a dark gray solid line,
while the best fit to \citet{asplund06} data (A(Li)=2.409+ 0.103[Fe/H]) is shown
by a dot-dashed gray line. The \citet{asplund06} Li abundances are increased
here by 0.04 dex to account for the known offset already mentioned in Sect.
\ref{compare_others}, and their metallicty is decreased by 0.2 dex to correspond
to the metallicity-scale offset detected by \citet{bonifacio07}. It is now even
more evident that the Spite plateau does not exist anymore at the lowest
metallicity, and is replaced by an increased spread of abundances, apparently
covering a roughly triangular region ending quite sharply at the plateau level.
This region appears here to be populated in a remarkably even manner; at any
probed metallicity some star remains at, or very close to, the Spite plateau
level, but many do not. The rather different slopes of the best-fit
relations in \citet{asplund06} and in this work appear to be the obvious
consequence of fitting two subsamples covering different  metallicity regimes.
This could provide also an explanation for the numerous claims, starting from
\citet{ryan99}, of a thin, but tilted Spite plateau. From this view, the
difference was produced simply because the tail of these samples had been falling in the
low-metallicity ``overdepletion zone'' as we have been able to
discern more clearly.

We are not aware of any theoretical explanation of this behavior. After the
measurements of the fluctuations of the CMB made it clear that there is a
``cosmological lithium problem'', i.e., the Li predicted by SBBN and the measured
baryonic density is too high with respect to the Spite plateau (by about 0.6\,
dex for our sample), there have been many theoretical attempts to provide
Li-depletion mechanisms that would reduce the primordial Li to the Spite
plateau value in a uniform way. Our observations now place anadditional constraint
on these models -- below a metallicity of about [Fe/H] = $-$2.5, they should
cause a dispersion in Li abundances and an overall lowering of A(Li).

If Li depletion from the WMAP-prescribed level were to happen in the stellar envelopes
of very metal-poor stars, the mechanism would have to be remarkably {\em metallicity
insensitive} to account for the thin, flat plateau observed between [Fe/H]=$-$2.5
and $-$1. And yet, the same phenomenon must become sharply {\em metallicity
sensitive} around and below [Fe/H]=$-$2.5, i.e., precisely where metallicity effects
on the atmospheric structure are expected to become vanishing small.

We are tempted to imagine that two different mechanisms may need to be invoked
to explain the production of the Spite plateau for stars with [Fe/H] $> -2.5$, and of the
low-metallicity dispersion for stars with [Fe/H] $< -2.5$. One could envision such
a two-step process as follows:
\begin{enumerate}
\item Metal-poor halo stars are always formed at the Spite plateau level, 
regardless of their metallicity. Whether the plateau represents the cosmological
Li abundance or is the result of some primordial uniform depletion taking place
{\em before} the star formation phase is immaterial in this context.
\item A second phenomenon, possibly related to atmospheric diffusion, becomes 
active around [Fe/H]=$-$2.5 and below, depleting Li further in the atmosphere of
EMP stars. This phenomenon, aside from the metallicity sensitivity, would exhibit
different star-to-star efficiency, being possibly dependent on additional
parameters, such as stellar rotation or \teff. Its efficiency must in any case
be higher for more metal-poor stars. 
\end{enumerate}
In this scenario, the ``primordial'' plateau would be preserved above
[Fe/H]$\sim$ $-$2.5, but below that metallicity, a systematic ``leakage'' of stars
towards lower A(Li) would take place, more effectively for more metal-poor stars, but
naturally scattered due to the sensitivity to parameters other than [Fe/H]. This 
scheme would have a number of advantages. First of all, it would naturally
explain our observations, ``mimicking'' a slope in A(Li) versus [Fe/H], but with
increased scatter at low [Fe/H]. It would also explain why, while the scatter in
A(Li) increases at low metallicities, not a single star in this metallicity
regime has been found to lie
above the Spite plateau level. It would then be consistent with a
small number of stars remaining close to the plateau at any metallicity
\citep[e.g., \object{CS 22876--032} A,][ filled magenta square in Fig.
\ref{alltogethernow}]{gonzalez08}; in these objects, the depletion process 
would be somehow inhibited. Finally, attributing the extra depletion to
atmospheric diffusion / settling would not require a physical ``conspiracy''
capable of producing exactly the same depletion level regardless of metallicity,
stellar rotation, gravity, or effective temperature, as is often invoked when
diffusion is used to explain the Spite plateau. 

The nature of what we refer to above as the ``second phenomenon'', the one responsible
for the departures from the Spite plateau below [Fe/H] = $-2.5$, is perhaps the most
intriguing. Above, we have proposed some kind of photospheric settling mechanism, but
one could as well envision a chemical evolution scenario, on the basis of some
gas pre-processing with Li depletion ({\it \`a la}} \citealt{piau06}) -- while
it may not be able to account for the entire WMAP-Spite plateau discrepancy,
this mechanism could easily account for the mild (0.2-0.4 dex) departure from the
plateau observed at lower metallicities. Moreover, this mechanism would
naturally produce a spread of abundances as a consequence of the local level of
gas pre-processing.

There are hints that the recently discovered ultra-faint dwarf galaxies
(uFdg) might have been the source of the bulk of the EMP stars now found in the halo of
the Milky Way \citep[][and references therein]{tolstoy09}. If this were indeed
the case, a sizeable fraction of our sample could have formed in uFdg systems, 
possibly more so for the most metal-poor objects. It has
been suggested \citep{komiya09} that the paucity
of stars below [Fe/H] = $-3.5$ may be due to the onset of self-pollution in the
primordial mini-halos when they started to merge to form larger structures, and
ultimately the halo. One could then envision that gas reprocessing could have
fairly significantly altered the Li abundances in the heavily dark-matter
dominated cores of these sub-halos, but become progressively negligible
when they merged, and their ISM mixed more and more completely with gas of
pristine Li abundance. Sub-plateau stars may then originate from star formation 
that occurred in the sub-halo cores, or while mixing with pristine gas that
progressively diluted the effect of the reprocessing.

In any case, the main drawback of this two-phenomena scenario is to leave the
WMAP-Spite plateau discrepancy unexplained. It also implies that any
depletion from WMAP-based primordial A(Li) should have taken place before the
currently observed stars formed. Non-standard primordial nucleosynthesis thus 
remains entirely viable, until the contrary is proven.

Looking retrospectively at the history of the abundances of light elements and
SBBN, one concludes that if the Spite plateau is the result of significant uniform
Li depletion among metal-poor stars, one faces a formidable case of cosmic
conspiracy. It is necessary to admit that the Li has been depleted
quite precisely to a level consistent with SBBN production and also
consistent with the abundances of the other measurable light elements. Had Li
been depleted to the level of A(Li)=1.8, that is, below the minimum allowed by
SBBN, the Spite plateau might not have been interpreted as being related to primordial Li, right
from the beginning.

A caveat is in order when looking at Fig. \ref{alltogethernow}: the \citet{asplund06} sample is biased, having been purposedly
selected not to include objects significantly deviating from the Spite plateau. In fact, stars with varying degrees
of Li depletion have long been known to exist \citep[e.g.,][]{charbonnel05}. On the other hand, 
{\em warm} (i.e., \teff$>$ 6000 K) dwarf stars deviating from the Spite Plateau appear to be quite rare
at [Fe/H]$>$-3. Again, our sample is limited towards higher metallicities, which prevents us from properly quantifying
this statement. A re-evaluation of the Li abundances above [Fe/H]=-2.5 is in order, using an analysis technique homogeneous with the present one, and
applied to a sample for which a well-known selection technique has been applied. 
It is plain to see that, had previous authors encountered Li abundance distributions as the one observed by us and \citet{aoki09}, the
concept of a lithium abundance plateau would have been short lived. Instead, there is a clear {\em perception} that mildly
Li depleted stars do exist among MP stars, but are a rare occurrence. 
We cannot easily identify any reason why all the previous studies should have been biased in a different way with respect to the present one,
but until a coherent work, based on a well defined selection criterion, is extended to higher metallicities, all 
we are left with is just this perception.

Two other lines of future investigation are suggested by the present results: ({\em i})
increase the sample of {\em hot} (\teff $\ge6250$\,K)) EMP stars ([Fe/H]$\le
-3.0$), and ({\em ii}) increase the number of cool (\teff $<6250$\,K) stars over the entire metallicity range. The present sample has seven {\em hot} EMP stars, five
of which have A(Li) below the Spite plateau. Even if ``standard'' ZAMS Li depletion
was setting in at higher than usual temperatures in EMP stars, these objects are unlikely
to be affected by it. Statistics is still weak in this metallicity regime, and that the primary component of
CS~22876--032 lies at the level of the Spite plateau, at a metallicity of [Fe/H]
= $-$3.6, clearly calls for a search for similar objects. 
At the same time, the cool end of our sample is presently too poorly populated to draw
any definitive conclusion about whether these stars are experiencing convective depletion, which again, calls
for an enlargement of the cool stars sample at low ([Fe/H]$<$-2.5) metallicities.

To help understand the phenomena involved (since there
could be several at work) that bring about the observed Li abundance pattern,
another important issue is to assign an accurate evolutionary status to each star,
that is, to confidently assess which stars are dwarfs, main-sequence turnoff, or
subgiants.  Our surface gravities are not sufficiently accurate for this purpose. The GAIA
satellite will provide accurate parallaxes for all of the presently studied
stars, thus allowing one to more clearly elucidate this problem.

\begin{acknowledgements}
Authors L. S., P. B., 
H.-G. L., N. B., and J. G.-H. 
acknowledge financial support from  
EU contract MEXT-CT-2004-014265 (CIFIST).
We acknowledge use of the
supercomputing center CINECA, which has granted us time to compute
part of the hydrodynamical models used in this investigation, through
the INAF-CINECA agreement 2006,2007.
T. C. B. acknowledges partial support for this work from grant 08-22648: Physics Frontier Center / Joint Institute for Nuclear Astrophysics.
This research has made use of the SIMBAD database, 
operated at CDS, Strasbourg, France, and of NASA's Astrophysics Data System.
This publication makes use of data products from the 
Two Micron All Sky Survey, which is a joint 
project of the University of Massachusetts and the 
Infrared Processing and Analysis Center/California 
Institute of Technology, funded by the National Aeronautics and 
Space Administration and the National Science Foundation. We also wish to thank K. Lind for making available
to us her code to interpolate the NLTE correction tables of \citet{lind09}.
\end{acknowledgements}

\appendix

\section{3D non-LTE treatment of lithium}
\label{3dnlte_calc}

The calculation of synthetic non-LTE \ion{Li}{I} $\lambda$~670.8\,nm 
line profiles from 3D hydrodynamical model atmospheres proceeds in two basic
steps. In the first step, the code {\tt NLTE3D} provides the departure 
coefficients $b_i(x,y,z,t) = n_i(x,y,z,t)/n_i^\ast(x,y,z,t)$, the ratio 
of non-LTE to LTE population number densities for each level $i$ of the 
\ion{Li}{i} model atom as a function of the geometrical position 
$(x,y,z)$ in the 3D model atmosphere, and time $(t)$ as sampled by a 
number of snapshots ($\approx 20$) selected to represent the characteristic 
temporal variation of the simulation. The departure coefficient of 
\ion{Li}{ii} is assumed to be $1$, since lithium is essentially fully
ionized in the stellar atmospheres of interest.

In the second step, the departure coefficients are fed into the
completely independent spectrum synthesis code
\linfor\footnote{\url{http://www.aip.de/~mst/Linfor3D/linfor_3D_manual.pdf}},
where they are used to compute the non-LTE line opacity and source function, 
and in turn the emergent \emph{intensity} profiles as a function of 
$(x,y,\theta,\phi,t)$, where the angles $\theta$ and $\phi$ specify
the orientation of the line-of-sight in polar spherical coordinates. 
Finally, the emergent mean \emph{flux} profile is obtained by horizontal, 
angular, and temporal averaging of the individual intensity profiles.

For this paper, we used an 8-level model atom of \ion{Li}{i} to solve
the statistical equilibrium equations, considering a total of 11 bound-bound
transitions. Details about the energy levels and line transitions are given 
in Tables \ref{tab:Liatom_levels} and \ref{tab:Liatom_transitions}.
The \emph{Einstein coefficients} $A_{j\,i}$ provided by the NIST database
are related to the \emph{Einstein coefficients} $B_{i\,j}$ by
\begin{equation}
A_{j\,i}=\frac{2h\nu_{i\,j}^3}{c^2}\,B_{j\,i} = 
\frac{2h\nu_{i\,j}^3}{c^2}\,\frac{g_i}{g_j}\,B_{i\,j}\, .
\end{equation}
The corresponding oscillator strength $f_{i\,j}$ is obtained from the relation
\begin{equation}
f_{i\,j}=A_{j\,i}\,\frac{g_j}{g_i}\,\frac{m_ec^3}{8\pi^2e^2\,\nu_{i\,j}^2} = 
B_{i\,j}\,\frac{m_ehc\,\nu_{i\,j}}{4\pi^2e^2}\, .
\end{equation}

%%%%%%%%%%%%%%%%%%%%%%%%%%%%%%%
\begin{table} 
\caption{Energy levels of the \ion{Li}{i} model atom.
Data are taken from the NIST database.
\label{tab:Liatom_levels}}
\centering
\begin{tabular}{ccccr}
\hline\noalign{\smallskip}
level & configuration & \multicolumn{2}{c}{energy} & statistical  \\
\#    &               & [Ryd]                & [eV] & weight  \\
\noalign{\smallskip}\hline\noalign{\smallskip}
1 &  2s &  0.0000000 & 0.00000 &  2 \\
2 &  2p &  0.1358136 & 1.84784 &  6 \\
3 &  3s &  0.2479204 & 3.37313 &  2 \\
4 &  3p &  0.2818128 & 3.83426 &  6 \\
5 &  3f &  0.2850726 & 3.87861 & 10 \\
6 &  4s &  0.3190534 & 4.34094 &  2 \\
7 &  4p &  0.3323350 & 4.52165 &  6 \\
8 &  4d &  0.3337369 & 4.54072 & 10 \\
\noalign{\smallskip}\hline\noalign{\smallskip}
\end{tabular}
\end{table}
%%%%%%%%%%%%%%%%%%%%%%%%%%%%%%%

%%%%%%%%%%%%%%%%%%%%%%%%%%%%%%%
\begin{table*} 
\caption{Bound-bound transitions of the \ion{Li}{i} model atom. 
Data are taken from the NIST database.
\label{tab:Liatom_transitions}}
\centering
\begin{tabular}{cccccccc}
\hline\noalign{\smallskip}
transition & lower & upper & transition    & $\lambda$    & \multicolumn{3}{c}{transition probability} \\
\#         & level & level & configuration & vacuum [\AA] & $A_{j\,i}$ & $B_{i\,j}$ & $f_{i\,j}$ \\
\noalign{\smallskip}\hline\noalign{\smallskip}
 1  &   1  &   2  &  2s-2p &  6709.7  & 3.72E+07 & 8.49E+10 & 7.532E-01 \\
 2  &   1  &   4  &  2s-3p &  3233.6  & 1.17E+06 & 2.99E+08 & 5.502E-03 \\
 3  &   1  &   7  &  2s-4p &  2742.0  & 1.42E+06 & 2.21E+08 & 4.802E-03 \\
 4  &   2  &   3  &  2p-3s &  8128.6  & 1.74E+07 & 7.84E+09 & 5.745E-02 \\
 5  &   2  &   5  &  2p-3d &  6105.3  & 5.11E+07 & 4.88E+10 & 4.759E-01 \\
 6  &   2  &   6  &  2p-4s &  4973.1  & 5.05E+06 & 5.21E+08 & 6.241E-03 \\
 7  &   2  &   8  &  2p-4d &  4604.1  & 1.64E+07 & 6.71E+09 & 8.687E-02 \\
 8  &   3  &   4  &  3s-3p & 26887.1  & 3.77E+06 & 5.53E+11 & 1.226E+00 \\
 9  &   3  &   7  &  3s-4p & 10795.1  & 3.69E+03 & 3.51E+07 & 1.934E-04 \\
10  &   4  &   8  &  3p-4d & 17550.0  & 4.89E+06 & 1.11E+11 & 3.763E-01 \\
11  &   5  &   7  &  3d-4p & 19281.0  & 3.31E+05 & 3.58E+09 & 1.107E-02 \\
\noalign{\smallskip}\hline\noalign{\smallskip}
\end{tabular}
\end{table*}
%%%%%%%%%%%%%%%%%%%%%%%%%%%%%%%

The computationally most expensive part of solving the non-LTE problem
is the calculation of the line-blanketed radiation field $J_\nu(x,y,z,t)$
at each grid point of the selected 3D models, which is needed to determine 
the photoionization rates for all atomic levels. This is done with a modified
version of the radiation transport routines that are used in the \cobold\
hydrodynamical simulations for computing the radiative energy exchange term 
$\vec{\nabla} \vec{F}_{\rm rad}(x,y,z)$. The solution of the radiative 
transfer equation is based on a Feautrier scheme applied to a set of long
characteristics. The continuous opacities used in this context are
computed with the routines {\tt IONDIS} \& {\tt OPALAM} from the Kiel stellar atmosphere 
package\footnote{\url{http://www.aip.de/~mst/Linfor3D/linfor.pdf}}. Line 
blanketing is taken into account by adding to the continuous opacity the 
opacity distribution functions (ODFs, `big division', $v_{\rm turb}=2$~km/s) 
of \citet{castelli03}, including the 
\ion{H}{i}\,-\,\ion{H}{}$^+$ and \ion{H}{i}\,-\,\ion{H}{i} quasi-molecular 
absorption near $1400$ and $1600$~\AA, respectively. There is a slight
inconsistency in the chemical composition adopted for the calculation of 
the opacities: the continuous opacities are based on the solar
abundances of \citet{sunabboasp}, while the ODFs rely on the solar 
composition of \citet{grevesse98}. To obtain the opacities for different
metallicities, the solar abundances were scaled by a global factor 
corresponding to the desired [M/H], with an enhancement of the 
$\alpha$-elements by 0.4~dex below [M/H]~$= -0.5$. A total of $600$ 
frequency points were used to obtain the $J_\nu$ between $\lambda$~$925$ and 
$19\,800$~\AA. We checked that treating continuous scattering as true 
absorption does not introduce any significant changes in the resulting 
departure coefficients.

Given the line-blanketed radiation field $J_\nu(x,y,z,t)$, the 
photoionization rate $P_{i\,\kappa}$ from level $i$ to the continuum
$\kappa$ is computed as
\begin{equation}
P_{i\,\kappa} = 
4\,\pi\,\int_{{\nu}_i}^\infty \frac{\alpha_i(\nu)\,J_\nu}{h\nu}\,
\mathrm{d}\nu\, \quad [s^{-1}] \quad ,
\end{equation}
where $\nu_i$ is the threshold photoionization frequency for level $i$,
and the photoionization cross-sections $\alpha_i(\nu)$ for all considered
atomic levels are taken from the TOPBASE Opacity Project on-line atomic
database at the `Centre de donn\'ees astronomiques de 
Strasbourg'\footnote{\url{http://cdsweb.u-strasbg.fr/topbase/topbase.html}}.
The photo-recombination rates $P_{\kappa\,i}$ from the continuum $\kappa$ 
to level $i$ are then given by
\begin{eqnarray}
P_{\kappa\,i} &=& 4\,\pi\,\int_{{\nu}_i}^\infty 
\frac{\alpha_i(\nu)\,B_\nu(T)}{h\nu}\,
\left(1-\exp\left\{-\frac{h\nu}{kT}\right\}\right)\,\mathrm{d}\nu \;\; + 
\nonumber \\
    &&  4\,\pi\,\int_{{\nu}_i}^\infty \frac{\alpha_i(\nu)\,J_\nu}{h\nu}\,
\exp\left\{-\frac{h\nu}{kT}\right\}\,\mathrm{d}\nu\, \quad [s^{-1}] \quad ,
\end{eqnarray}
$B_\nu(T)$ denoting the Kirchhoff-Planck function at local temperature $T$ 
and frequency $\nu$.

Cross-sections for the collisional ionization and excitation by electrons 
are computed according to the prescriptions of \citet{seaton1962} and 
\citet{vregemorter1962}, respectively, as given by \citet{allen1976} and 
\citet{allen2000}. The oscillator strengths $f_{i\,j}$ from Table
\ref{tab:Liatom_transitions} are needed to calculate the
collisional excitation cross-sections.
Collisional ionization by neutral hydrogen via the charge transfer reaction 
H($1s$) + Li($nl$) $\rightarrow$ Li$^+$($1s^2$) + H$^-$, and the reverse
process H$^-$ + Li$^+$($1s^2$) $\rightarrow$ H($1s$) + Li($nl$), 
are treated according to \citet{barklem2003} for the first 7 levels. 
Collisional excitation by neutral hydrogen is ignored, as it was found
to be unimportant for thermalizing \ion{Li}{i} by \citet{barklem2003}.

The departure coefficients $b_i(x,y,z)$ are finally obtained by solving 
the statistical equilibrium equations locally at each grid point $(x,y,z)$.
The profile-averaged radiation field at the line transitions, 
$\overline{J}(\nu_{i\,j})$, determines the radiative excitation rates
\begin{equation}
R_{i\,j}= B_{i\,j}\, \overline{J}(\nu_{i\,j}) \equiv
B_{i\,j}\,\int_{{\rm line}\; i\,j} \phi_\nu\,J_\nu\, \mathrm{d}\nu\, 
\quad [s^{-1}] \quad ,
\end{equation}
and the radiative de-excitation rates
\begin{equation}
R_{j\,i}=\frac{g_j}{g_i}\,B_{j\,i}\,\left(\frac{2h\nu_{i\,j}^3}{c^2} + 
\overline{J}(\nu_{i\,j})\right)\, \quad [s^{-1}] \quad ,
\end{equation}
where the statistical weights of lower and upper level, 
$g_i$ and $g_j$, and the \emph{Einstein coefficients} $B_{i\,j}$ are
taken from Table\,\ref{tab:Liatom_transitions}. For the line profile
$\phi_\nu$ we assume a purely Gaussian distribution (identical for
absorption and emission) with a line width that corresponds to the 
local thermal Doppler velocity plus a microturbulence of $1.5$~km/s. 
The Doppler shift caused by the line-of-sight component of the hydrodynamical
velocity field is ignored in the present computation. For all transitions except for
the resonance line, $\overline{J}(\nu_{i\,j})$ was replaced by the 
mean continuum intensity $J_\nu$. For the resonance line (transition \# 1),
the line opacity was taken into account for the computation of
$\overline{J}(\nu_{i\,j})$, assuming an $f$-value of $0.48975$ 
(about 2/3 of the value given in Table\,\ref{tab:Liatom_transitions} 
to represent only the main component of the doublet), together with a 
typical lithium abundance of A(Li)=$2.2$.

Since $\overline{J}(\nu_{i\,j})$ is a non-local quantity that depends in 
turn on the $b_i$, a $\Lambda$-iteration is employed to obtain a consistent 
solution. Fortunately, the $\Lambda$ iteration converges very rapidly, 
because even the \ion{Li}{i} resonance line affects the radiation field 
only marginally, while all the other lines are very weak.
Typically, three iterations are sufficient to achieve convergence.
 
Finally, the departure coefficients $b_i$ are used in the line
formation code \linfor\ to compute
the non-LTE line opacity
\begin{equation}
\kappa_{\rm line}{\rm (non-LTE)} =
\frac{b_{\rm low}-b_{\rm up}\,
\exp\left\{-\frac{h\nu}{kT}\right\}}
{1-\exp\left\{-\frac{h\nu}{kT}\right\}}\,\kappa_{\rm line}{\rm (LTE)}\, ,
\end{equation}
and the line source function
\begin{equation}
S_{\nu,{\rm line}}{\rm (non-LTE)} = b_{\rm up}\,
\frac{1-\exp\left\{-\frac{h\nu}{kT}\right\}}
{b_{\rm low}-b_{\rm up}\,\exp\left\{-\frac{h\nu}{kT}\right\}}\,B_\nu\, ,
\end{equation}
where $b_{\rm low}$ and $b_{\rm up}$ are the departure coefficients of
lower and upper level, respectively, and $B_\nu$ is again the Kirchhoff-Planck 
function.

\section{Analytical fit to 3D NLTE Li abundance as a function of stellar parameters and EW}
\label{fitappendix}

\subsection{Introduction}

It is often convenient to have tabular data on an irregular grid $y=y(x)$
condensed into a single function (even with many fit parameters), instead of
having to interpolate between the original data points. Choosing a fitting function
that is linear in all parameters (e.g., a sum of polynomials as $A + B x + C
x^2$ or sine/cosine functions) results in a comparatively simple linear problem
for finding these parameters, usually with one unique solution. However, the
result might not have the desired extrapolation behavior, might be ``too
wiggly'' between the data points, or produce no good fit at all.

A more effective approach might then be to start with a properly crafted
function, possibly non-linear, and having fitting parameters ``within'' the
terms as in $A \exp(B + C x)$. In this case, an iterative process is necessary
to determine these parameters. This iteration requires starting values, which will
affect the convergence of the process and also the solution if one is found: in this case,
the scheme might diverge or converge to different local minima, depending on the
starting point. However, this last issue is hardly a problem for our application
-- we are content with a good fit, and do not necessarily need to achieve the closest
one. We continue, however, to have the problem of constructing the fitting function and choosing the
initial values for the parameters.

\subsection{Method}

We used a scheme for recursive term substitution, written in IDL, that attempted
not only to find optimum fitting parameters for a given function, but also to
automatically determine the optimum functional form itself. The latter was
chosen from a set of candidates given by a list of terms and some construction
rules. The usual starting point was a constant ``function'' with one parameter,
say $A$, the best ``fit'' for which is the mean of $y$. In each recursion step,
all parameters were successively substituted with all of the terms from a list
(with possible restrictions), for instance
\begin{displaymath}
  A \rightarrow A_0 + A_1 x,\,\,
\end{displaymath}
or
\begin{displaymath}
   A \rightarrow A_0 + A_1 \exp(A_2 x).
\end{displaymath}
The initial values of ($A_0$, $A_1$, $A_2$) were derived from $A$ by e.g., assuming
  $A_0=A, A_1=0.1\times A, A_2=0$.
Applying a modified version of the IDL function {\tt curvefit}
(requiring also partial derivatives) provided the optimum parameter set for the given function and start values.

This process was repeated with a number of random variations
to approach an optimum solution for the given function.
The final solution was the globally best fit among all candidate functions and parameters.

The main control parameters are the list of candidate terms and the recursion depth.
For the current application, we search a 4D function
\begin{displaymath}
   y = y(x_0, x_1, x_2, x_3)
\end{displaymath}
where $(x_0, x_1, x_2, x_3)=($ $\log$(EW), $\log$\teff, \glog, [Fe/H]$)$, or, for the inverse problem,
$($ A(Li), $\log$\teff, \glog, [Fe/H]$)$.
We start with
\begin{displaymath}
  A_0 + \log_{10}[1 - \exp(- 10^{(X_0-A_1)})]
  \end{displaymath}
to determine the equivalent width as a function of abundance,
based on a simple absorption model for a box-shaped line profile.
In the following recursion steps, we replace the parameters with linear terms
\begin{displaymath}
  A \rightarrow A_0 + A_1*x_1 + A_2 x_2 + A_3 x_3,
\end{displaymath}
which finally results in higher-order polynomials instead of the initial coefficients.
In this step, the variation in the COG with the stellar parameters is taken into
account. In the process of applying the term substitution, we scanned through several hundred
possible functional forms. We decided the most suitable one on the basis of
simplicity, and its ability to represent the numerical data with sufficient accuracy.

As described in Sect. \ref{3dnlte}, synthetic grids were computed for three cases,
3D NLTE, 1D {\tt LHD} NLTE, and 1D {\tt LHD} LTE. This in principle leads to 6 fitting functions being computed,
when both the EW$\rightarrow$A(Li) and A(Li)$\rightarrow$EW forms are required.
We decided to force all three functions in either sense to have the same
form, differing only by their parameters. We thus first searched an optimum
function for one case (A(Li)$\rightarrow$1D {\tt LHD} NLTE turned out to be the
optimum choice) and then varied only the parameters (and not the functional
form) for the other two cases. The final inversion of the function was performed by
hand.

\subsection{Best-fitting functions}

Best-fitting functional forms were produced both to derive A(Li) from EW
and to derive EW from A(Li), the latter being useful, e.g., for the preparation of
observations. Analytical fit formulae for both cases are presented in Eq.
\ref{fitform}; fitting coefficients are listed in Table \ref{coeffit}. Three
sets of coefficients are listed. The 3D NLTE set is used to fit computations where
\cobold\ 3D hydrodynamical models are used in association with a 3D NLTE
time-dependent spectrosynthesis. The 1D NLTE and LTE cases refer instead to
computations performed by using {\tt LHD} one-dimensional models, with
spectrosynthesis performed with or without the inclusion of NLTE effects, based
on the same physics and model atoms as for the 3D NLTE case:

\begin{equation}
\begin{array}{l}
Q_1  =  A_8 + A_{11}\log T_{\rm eff} + A_{12} \log g + A_{13} \mathrm [Fe/H] \\
Q_2  =  A_6 + Q_1 \log T_{\rm eff} +A_9 \log g + A_{10} \mathrm [Fe/H] \\
Q_3  =  A_2 + A_5 \log T_{\rm eff} + Q_2 \log g + A_7 \mathrm [Fe/H] \\
Q_4  =  A_1+Q_3 \log T_{\rm eff} +A_3 \log g + A_4\mathrm [Fe/H]\\
\\
A(Li)  =  A_1 + Q_3 \log T_{\rm eff} + A_3 \log g + A_4 \mathrm [Fe/H]\\
~~~~~~~~~~~~~~~~~~~~~~~~~~~~~~+\log[-\ln(1-EW 10^{-A_0}) ]\\
\\
\log EW  =  A_0 + \log [ ~| 1 + \exp (-10^{A(Li)-Q_4})|~ ].
\end{array}
\label{fitform}
\end{equation}

\begin{table}
\caption{ Coefficients for the analytical fit of  EW $\rightarrow$ A(Li) and A(Li) $\rightarrow$ EW \label{coeffit}}
\begin{center}
\begin{tabular}{lr@{.}lr@{.}lr@{.}l}
\hline
         & \multicolumn{2}{c}{\bf 3D NLTE}   & \multicolumn{2}{c}{\bf 1D NLTE}   & \multicolumn{2}{c}{\bf 1D LTE}   \\
\hline
A$_{0}$  &  2&1744416E+00  &  2&2011840E+00  &  2&1694977E+00 \\
A$_{1}$  &  3&9685178E+02  & -2&3708574E+02  & -3&1329205E+02 \\
A$_{2}$  & -2&1920459E+02  &  9&9754471E+01  &  1&5692142E+02 \\
A$_{3}$  & -9&8448749E+02  &  1&5211893E+02  &  1&9027467E+03 \\
A$_{4}$  &  2&4222436E+00  & -7&0910416E+00  &  8&0236683E+00 \\
A$_{5}$  &  3&0470810E+01  & -9&5952034E+00  & -1&9364346E+01 \\
A$_{6}$  &  7&4822784E+02  & -1&0479124E+02  & -1&4933164E+03 \\
A$_{7}$  & -6&0743892E-01  &  1&8554645E+00  & -2&1075335E+00 \\
A$_{8}$  & -1&8920995E+02  &  2&5357561E+01  &  3&9087711E+02 \\
A$_{9}$  & -7&9977289E-02  & -7&5381881E-01  & -2&7466467E-01 \\
A$_{10}$ & -3&2394665E-01  &  3&8146901E-01  & -4&9764225E-01 \\
A$_{11}$ &  1&5911137E+01  & -2&1918788E+00  & -3&4126404E+01 \\
A$_{12}$ &  2&3078753E-02  &  1&9988796E-01  &  7&3497586E-02 \\
A$_{13}$ &  8&3344564E-02  & -9&9645615E-02  &  1&3085939E-01 \\
\hline
\end{tabular}
\end{center}
\end{table}

Units are expected to be K for temperature, cm s$^{-2}$ for \glog, and m\AA\ for
EW. As for any fit, extrapolation reliability is difficult to assess. Owing to the high
computational cost of 3D-atmosphere model calculations, we limited our grid to the
currently available models, which obliged us to extrapolate towards high
gravities by 0.4 dex and towards low metallicities by about 0.6 dex. Both
extrapolations should be quite safe, since the 670.8 nm Li doublet should be
quite insensitive to both parameters. Because of the vanishing line
opacity, atmosphere models are scarcely sensitive to metallicity variations
below [Fe/H] = $-3$. Those who might wish to employ the presented formulae to derive Li
abundances are nevertheless advised to use caution when extrapolating,
especially in \teff\ and EW. In particular, the saturation part of the COG is
almost unsampled, and one cannot expect the present fit to reproduce it
properly.  Fitting to EW significantly above 100 pm is therefore not advisable.

\end{document}